\documentclass[aps,prb,twocolumn, showpacs]{revtex4}
\usepackage{graphicx}
\usepackage{dcolumn}
\usepackage{amsmath}
\newcommand{\gamnas}{${\rm Ga}_{\rm 1-x}{\rm Mn}_{\rm x}{\rm As}$\;}
\newcommand{\be}{\begin{equation}}
\newcommand{\ee}{\end{equation}}
\newcommand{\bea}{\begin{eqnarray}}
\newcommand{\eea}{\end{eqnarray}}
\newcommand{\nl}{\nonumber \\}

\newcommand{\mopt}{m_{\rm opt}}
\newcommand{\mstar}{m^\star}

\newcommand{\expect}[1]{\large\langle #1 \large\rangle}

\begin{document}

\title{Theory of optical conductivity for dilute \gamnas}
\author{C. P. Moca$^{1,2}$, G. Zar\' and$^1$, M. Berciu$^3$ }
\affiliation{
$^1$Department of Theoretical Physics, Budapest University of Technology and Economics, Budapest, Hungary\\
$^2$Department of Physics, University of Oradea, 410087 Oradea, Romania \\
$^3$Department of Physics and Astronomy, University of British
  Columbia, Vancouver, British Columbia, Canada V6T1Z1
}

\date{\today}
\begin{abstract}
 We construct a semi-microscopic theory, to describe the optical
 conductivity of \gamnas in the dilute limit, $x\sim 1 \;\%$. We
 construct an effective Hamiltonian that captures inside-impurity band
 optical transitions as well as transitions between the valence band
 and the impurity band. All parameters of the Hamiltonian are computed
 from microscopic variational calculations. We find a metal-insulator
 transition within the impurity band in the concentration range,
 $x\sim 0.2 -0.3\; \%$ for uncompensated and $x\sim 1 -3\; \%$ for
 compensated samples, in good agreement with the experiments. We find
 an optical mass $\mopt \approx m_e$, which is almost independent of
 the impurity concentration excepting  in the vicinity of the
 metal-insulator transition, where it reaches values as large as
 $\mopt \approx 10\;m_e$.  We also reproduce a mid-infrared peak at
 $\hbar \omega\approx 200 \;{\rm meV}$, which redshifts upon doping,
 in quantitative agreement with the experiments.
\end{abstract}

\maketitle

\section{Introduction}

The diluted magnetic semiconductor (DMS) Ga$_{1-x}$Mn$_{x}$As has
emerged as one of the most promising materials due to its potential
applications in spin-based technology. This material is, however,
equally interesting from the point of view of fundamental research for
its unique properties and the rich physics it displays; \gamnas is a
strongly disordered ferromagnet, where the interplay of
ferromagnetism, localization, magnetic
fluctuations and the presence of strong spin-orbit coupling lead to
many interesting properties such as a strong
magnetoresistance,\cite{Ohno,Zarand} resistivity
anomalies,\cite{Zarand, vonOppen, Novac, Brey} or the presence of a large
anomalous Hall effect.\cite{MacDonald}

Although \gamnas has been the subject of intense theoretical and
experimental investigation, even its most basic properties are still
debated. One of these fundamental and unresolved issues is the
existence or non-existence of an impurity band in this material. There
is a general consensus that for very small magnetic concentrations,
$x<1\%$, \gamnas is described in terms of an impurity band. For higher
concentrations, however, there is no general consensus yet: On one
hand, essentially all spectroscopic measurements such as
ARPES,\cite{Okabayashi} STM,\cite{STM} optical conductivity
data,\cite{Singley1,Singley2,Burch1,Burch2} and ellipsometry
\cite{Linnarsson} seem to favor the presence of an impurity band even
up to moderate concentrations ($x\approx 3-5 \%$), and even high
quality samples with a high Curie temperature and a clearly metallic
behavior have surprisingly small values of $k_Fl\sim
1$.\cite{Zarand,Burch2} On the other hand, many properties of these
materials can be even quantitatively understood in terms of a
disordered valence band picture, and, in fact, from a theoretical
point of view, it is hard to understand how an impurity band could
survive up to concentrations as high as $x\approx
5\%$.\cite{Jungwirth3}

In the present paper we shall not attempt to resolve this discussion,
rather, we would like to focus exclusively on the {\em optical
  conductivity}.  Typical optical conductivity measurements in \gamnas
display two rather remarkable features: $(i)$ a mid infrared resonance
peak at approximately $200\; meV$ that {\em redshifts} with increasing
hole concentration, $p$; $(ii)$ a large optical mass of the order
$\mopt \sim 0.7 \div 1.5 m_e$, implying a mobility which is orders of
magnitude smaller than the one of ${\rm GaAs}$ doped with similar
concentrations of non-magnetic impurities.\cite{Liu}

In this paper we aim at developing a semi-microscopic theory for the
optical conductivity of \gamnas in the {\em very dilute limit}, $x\sim
1 \%$.  There are several theoretical studies of the optical
conductivity of \gamnas by now.\cite{Yang, Turek, Hwang, Hoang, Alvarez}
Maybe the most realistic calculations have been carried out in
Refs.[\onlinecite{Yang}] and [\onlinecite{Turek}], however, the
results of these works are somewhat conflicting: In
Ref.~[\onlinecite{Yang}] a disordered valence band described in terms
of the Luttinger model has been studied. In these calculations, rather
surprisingly, the mid-infrared peak appeared even in a simplified
parabolic model, where a completely isotropical valence band mass was
assumed and the spin-orbit split valence band is completely
ignored. Therefore the authors concluded that the mid-infrared peak is
not due to transitions to the spin-orbit split valence band, rather,
they interpreted it as a result of strong localization and
interference effects.  In another, very thorough
calculation\cite{Turek}, realistic tight binding models have been
used, but the results were conflicting in that features associated
with an impurity band have or have not appeared depending on the
particular tight binding scheme applied.

Unfortunately, none of these previously applied methods is appropriate
to study accurately the dilute limit, $x \sim 1\%$. This small
concentration limit is of particular interest, since the
metal-insulator transition takes place at concentrations as low as
$x\approx 0.5\;\%$,\cite{Blakemore} and furthermore, it is in this
limit where features associated with the impurity band should be
present.  In the present paper, we shall follow the lines of
Ref.~[\onlinecite{Fiete}] and develop a semi-microscopical approach to
capture the physics of this dilute limit. At the microscopic level, we
describe the valence band holes as spin $F=3/2$ carriers, which
interact with the external electromagnetic vector potential  
and the Mn ions, as described by the following
Hamiltonian: \bea H&=& \sum_i\frac 1{2 m_0} ({\bf p}_i-q {\bf A}({\bf
  r}_i))^2 + \sum_{i,m} V({\bf r}_i-{\bf R}_m) \nl &+& \sum_{i,m} J
\;\delta({\bf r}_i-{\bf R}_m) \; {\bf F}_i \cdot {\bf S}_m \;.
\label{eq:micro}
\eea Here ${\bf r}_i$ and ${\bf p}_i$ ($i=1,\dots, N_h$) denote the
valence hole coordinates and momenta, and the Mn atoms are located at
random positions, ${\bf R}_m$ ($i=1,\dots, N_{\rm Mn}$).  In the first
term we assume an isotropic effective mass $m_0 = 0.56\, m_e$. This is
a somewhat uncontrolled approximation. However, according to the
results of Ref.~[\onlinecite{Yang}], inclusion of a more realistic
band structure influences only slightly the optical spectrum for
frequencies $\hbar \omega< 800 {\rm meV}$, we shall therefore ignore
it here. The second term describes the random potential created by the
Mn atoms, and it incorporates the Coulomb attraction as well as the
central cell correction.\cite{central_cell} Finally, the last term
describes the local exchange interaction between the manganese
magnetic moments, ${\bf S}_m$ and the spin of hole $i$, ${\bf
  F}_i$. The parameters of the 'microscopic' Hamiltonian
Eq.~\eqref{eq:micro} are all well-known.\cite{Jungwirth1}

In our work we use the Hamiltonian \eqref{eq:micro} as a starting
point, and derive a microscopic effective Hamiltonian from it that
captures the impurity band physics in the very dilute limit, but also
accounts for impurity band-valence band transitions (for details see
Section~\ref{sec:models}).  This effective Hamiltonian can then be
used to compute the optical conductivity using field theoretical
methods, as detailed in Section~\ref{sec:opt_cond}. With this
'multiscale approach', we are then able to determine the optical
properties of \gamnas accurately in the small concentration regime,
while keeping $N_h\approx 200$.

\begin{figure}[t]
\centering
\includegraphics[width=3.0in,clip]{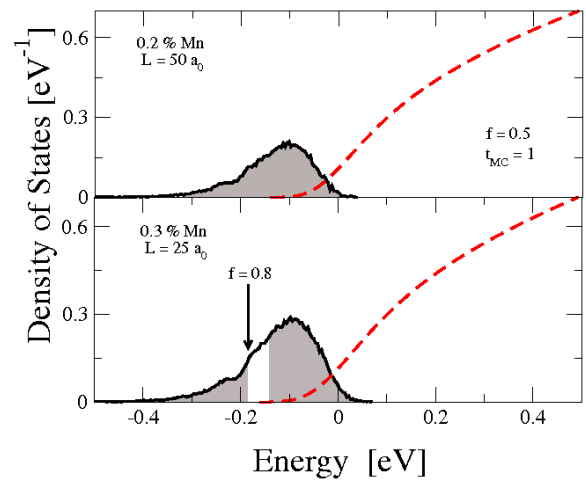}
\caption{Density of states on for $x=0.2\%$ and $x=0.3\%$. Both the
  impurity band and valence band contributions are shown. At these
  concentrations the impurity band is clearly separated from the
  valence band. The gray area indicates the regime of localized
  states, while delocalized states are shown as white regions.  For
  these calculations we used a hole fraction $f=0.5$. The system size
  is measured in the size of the unit cell, $a_0$.  }
\label{fig:MIT}
\end{figure}
One of the key elements of our approach is the assumption that, for
these small concentrations, the {\em Fermi level lies inside the
  impurity band}.  This is indeed evidenced by transport
measurements,\cite{Blakemore} which clearly show that the activation
energy (associated with localized impurity states to valence band
transitions) remains finite as one approaches the metal insulator
transition.\cite{Blakemore} This implies that the metal-insulator
transition takes place {\em within the impurity band},
and one must be able to capture the properties of small concentration
metallic \gamnas samples using our impurity band-based method.

 Indeed, our results also support this picture; a detailed analysis of
 the hole states reveals a metal-insulator transition at about
 $x\approx0.2\div 0.3\;\%$, where delocalized states appear {\em in
   the middle of the impurity band}. As shown in Fig.~\ref{fig:MIT},
 this is approximately the critical concentration for uncompensated
 samples.  The Anderson transition thus takes place {\em inside the
   impurity band}, in agreement with the conclusions of
 Ref.~\onlinecite{Fiete}, and also in good agreement with activation
 energy values observed for samples grown at high
 temperature.\cite{Blakemore1,Moriya} Typically, however, \gamnas
 samples grown under non-equilibrium conditions are {\em compensated}
 even after annealing: in addition to the substitutional Mn ions (of
 concentration $x_{S}$) there is also a finite concentration $x_{I}$
 of interstitial Mn ions, which behave as double donors, and are also
 believed to make a fraction of the Mn ions inactive by simply binding
 to them.\cite{Awschalom1} As a result, the effective concentration of
 ``active'' Mn ions is reduced to $x\equiv x_{\rm eff}\approx
 x_{S}-x_I $ and the concentration of holes is also suppressed
 compared to $x_{\rm eff}$ by the hole fraction $f\approx(x_{S}-2
 x_I)/(x_{S}-x_I)$. Remarkably, even if only 20$\%$ of the Mn ions
 goes to interstitial positions, that reduces the effective
 concentration to 60$\%$ percent of the total Mn concentration and
 amounts in a hole fraction $f \approx 0.66\%$. Unannealed samples
 tend to have even larger interstitial Mn concentrations and thus much
 smaller hole fractions.\cite{Awschalom1,rutherford,Jungwirth1} As a
 result, depending on the precise annealing protocol, ferromagnetic
 \gamnas samples (grown under non-equilibrium conditions) tend to show
 the phase transition at higher concentrations.  According to our
 calculations, for $x_I/x_{\rm total} = 30 \%$ ($f=0.3$) the metal
 insulator transition takes place at about $x_S\approx 3.2 \;\%$,
 while for $x_I/x_{\rm total} = 23 \%$ ($f=0.55$) the transition
 occurs at $x_S\approx 0.7 \;\%$. These values are consistent with the
 experimental data.\cite{Oiwa97, Blakemore1,Matsukara, Esch, Shimizu}
 In the rest of the paper, we shall not make a difference between
 substitutional and interstitial Mn ions, rather, by the symbol $x$ we
 shall refer to the concentration of {\em active / effective} Mn ions,
 which participate in the formation of the ferromagnetic state, and
 $f$ will be the corresponding hole fraction.

We also find that the optical mass increases to very large values
$\mopt \sim 10\; m_e$ at the critical concentration, where the DC
conductance vanishes. However, apart from this, the optical spectrum
changes rather continuously. For larger Mn concentrations, $x>2-3~\%$,
we obtain an optical mass $\mopt \approx m_e$, which is almost
independent of the hole concentration $p$, and is in surprisingly good
agreement with the experimental values.\cite{Burch1}

Our approach is designed to work in the limit of small concentrations,
and it should break down for large concentrations. Where exactly this
breakdown takes place, is not quite clear. Our calculations show that
the impurity band is {\em not merged} with the valence band for
concentrations as large as $x\approx 3\%$. Furthermore, the impurity
band picture may make sense for even larger concentrations if the
Fermi level is inside the gap (as indicated by ARPES
measurements),\cite{Okabayashi} where states are expected to exhibit a
stronger impurity state character. We are not able to convincingly
determine the upper limit of our approach. However, if we blindly
extend it to the regime $x>3\%$, without justifying their
applicability, rather surprisingly, our calculations {\em
  qualitatively} as well as {\em quantitatively} reproduce all
important features of the experiments.

\begin{figure}[t]
\centering \includegraphics[width=3.0in,clip]{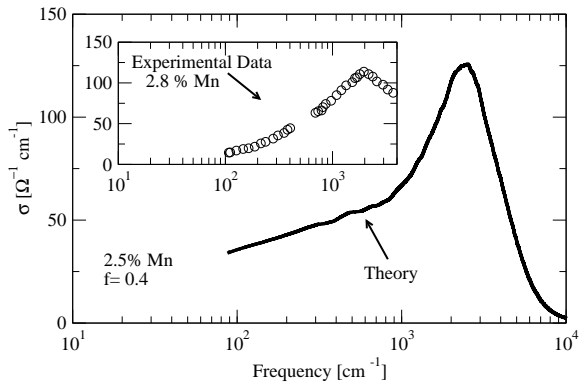}
\caption{Comparison between our theory and experimental data. The
  optical conductivity data presented in the inset were extracted from
  Ref.~[\onlinecite{Burch1}].}
\label{fig:comparison}
\end{figure}
One of the typical optical conductivity results is displayed in
Fig.~\ref{fig:comparison}, where, for comparison, we also show some
typical experimental data.\cite{Burch1}  In our calculations the Drude
contribution originates entirely from carriers residing within the
impurity band, and the 'Drude peak' appears as a plateau at smaller
frequencies, just as in the optical conductivity experiments. The
mid-infrared peak, on the other hand, is due to transitions from the
impurity to the valence band. Notice that the position as well as the
overall size of the signal are quantitatively reproduced.

\begin{figure}[t]
\centering
\includegraphics[width=3.0in,clip]{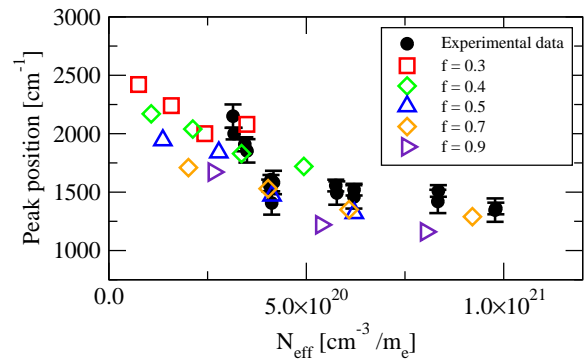}
\caption{Mid-infrared peak position extracted from the optical
  conductivity curves vs.the effective optical spectral weight,
  $N_{\rm eff}$ (defined through Eq.~\eqref{eq:N_eff}).  Open symbols
  represent theoretical results while the solid dots are the
  experimental data extracted from
  Ref. [\onlinecite{Burch1}]. Calculations for different hole
  fractions $f$ and concentrations $x$ ($x= 1\div 5\% $) fall onto a single
  curve.}
\label{fig:peak_positions}
\end{figure}

Having the Fermi level within the impurity band naturally explains the
redshift of this resonance with increasing hole concentration.  In
Fig.~\ref{fig:peak_positions} we show the computed peak position as a
function of the effective optical spectral weight, $N_{\rm eff}$
(defined through Eq.~\eqref{eq:N_eff}) for a variety of concentrations
$x$ and hole fractions, $f=N_h/N_{\rm Mn}$. Our calculations show
excellent agreement with the experimental results of
Ref.~\onlinecite{Burch1}.

The paper is organized as follows. First, in Section~\ref{sec:models}
we outline the calculations that lead from the microscopic Hamiltonian
Eq.~\eqref{eq:micro} to the effective Hamiltonians used. Some of the
details of these rather technical variational calculations are given
in appendices \ref{app:1Mn} and \ref{append:2Mn}. Sec.~\ref{sec:opt_cond}
provides the basic expressions for the optical conductivity within a
mean field approach, while the results of our numerical calculations
are given in Sec.~\ref{sec:mean_field_res}. In
Sec.~\ref{sec:fluctuations} we show how to incorporate the effects of
magnetic fluctuations, and finally, we conclude in
Sec.~\ref{sec:conclusions}.

\section {Theoretical framework}
\label{sec:models}

With current day computer technology, it is  essentially
impossible to treat
the Hamiltonian of Eq.~ \eqref{eq:micro} accurately for small Mn
concentrations, $x\sim 1\%$. We therefore describe this regime in
terms of an effective Hamiltonian, which we derive from
Eq.~\eqref{eq:micro}, and which consists of an impurity band part
and a valence band part. The first part of the present Section
focusses on this mapping.  However, to compute the optical
conductivity, we also need to determine how the electromagnetic field
in Eq.~\eqref{eq:micro} couples to impurity states within the
effective Hamiltonian. This is discussed in the second part of the
Section. In the main body of the text we only outline and summarize
the most important steps. Details on these simple but rather lengthy
calculations are given in Appendices \ref{app:1Mn}-\ref{append:2Mn},
and are provided as supplementary information.

 \subsection{Effective Hamiltonian}
\label{sub:EffectiveHamiltonian}

\subsubsection{Impurity band Hamiltonian}
Since for very small concentrations the Fermi energy is in the
impurity band, we first need to describe the impurity states. In the
small concentration limit, hole wave functions are localized at the Mn
sites, providing a strong and attractive potential for them. Therefore
the impurity band can be described using a tight binding Hamiltonian
of the form,\cite{Berciu,Fiete}
\begin{eqnarray}
H_{\rm imp}&=&-\sum\limits_{i,j,\alpha }t_{ij}\;c_{i\alpha }^{\dagger
}c_{j\alpha }
\label{eq:H_imp}
\\ &+&\sum\limits_{i,\alpha }E_i\; c_{i\alpha }^{\dagger }c_{i\alpha }
+ G \sum\limits_{i,\alpha,\beta }{\bf S}_{i} \cdot c_{i\alpha
}^{\dagger } {\bf F}_{\alpha \beta} c_{i\beta }\; .  \nonumber
\end{eqnarray}
Here $c_{i\alpha }^{\dagger }$ creates a bound hole at Mn position
$\mathbf{R}_{i} $ in a spin state $\left| F=3/2,F_z=\alpha
\right\rangle$, and $t_{ij}$ denotes the hopping between Mn sites
$\mathbf{R}_{i}$ and $\mathbf{R}_{j}$. Since the hole mass is
isotropic in our microscopic Hamiltonian, the hopping is independent
of the hole spin $\alpha$.  \footnote{A careful calculation using the
  full six band Hamiltonian shows that the hopping is indeed fairly
  isotropic as long as the hopping distance is less than about $R\sim
  15 \AA$ .\cite{Konig}} The on-site energy $E_{i}$
contains the binding energy of the hole, $E_0= -110 \;{\rm meV}$, but
it also accounts for the Coulomb and kinetic energy shifts generated
by neighboring Mn ions. Finally, the last term describes the local
antiferromagnetic exchange with the core Mn spin ${\bf S}_{i}$.

\begin{figure}[t]
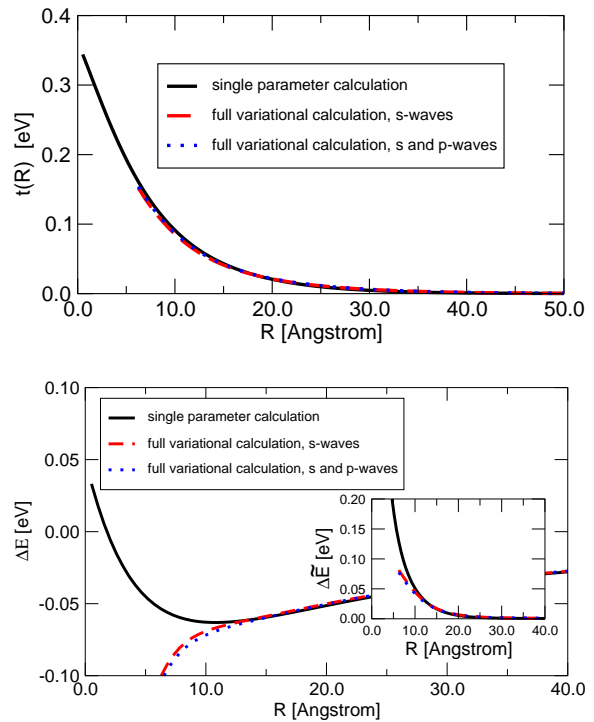

\centerline{\includegraphics[width=3.0in,clip]{comparison_t}}\vspace*{3ex}
\centerline{\includegraphics[width=3.0in,clip]{comparison_delta_e}}\vspace*{3ex}
\caption{Top: Variationally computed hopping parameter $t$ as function
  of the relative distance between the Mn sites, $R$. Bottom: The
  energy shift $\Delta E (R)$.  The inset shows the correction
  ${\Delta \tilde E} (R)$ to the classical Coulomb shift.}
\label{comparison_t_fig}
\label{comparison_delta_e_fig}
\end{figure}

The parameters appearing in Eq.~\eqref{eq:H_imp} can be determined
from experiments and from the microscopic Hamiltonian,
Eq.~\eqref{eq:micro}.  The coupling $G$  is known
from EPR experiments to be $G\approx 5 {\rm meV}$ for a single Mn
impurity.\cite{Ohno_coupling, Linnarsson} However, the hopping
parameters and the on site energy, $E_i $ need to be computed from
Eq.~\eqref{eq:micro}. Similar to Ref.~\onlinecite{Fiete}, we
determined them from a variational calculation of the molecular
orbitals for an Mn$_2$ 'molecule', described by the Hamiltonian:
\begin{equation}
H_{\rm 2site}= -\frac{\gamma }{2}\nabla^{2}-\frac{1}{\epsilon
  r_{1}}-\frac{1}{ \epsilon r_{2}} + V_{cc}(r_{1})+V_{cc}(r_{2})\;,
 \label{eq:2site}
\end{equation}
where we used atomic units, $r_{1,2} = |{\bf r} - {\bf R}_{1,2}|$,
$\epsilon = 12.65$ is the dielectric constant of ${\rm GaAs}$, and $\gamma =
m_e/m_0 = 1.782$. The so-called central cell correction $V_{cc}$
accounts for the local interaction at the Mn core, and is given
by\cite{cc_correction}
\begin{equation}
V(\mathbf{r})=V_{cc}(r)=-V_{0}\;e^{-r/r_{0}}\;.
\label{eq:V_cc}
\end{equation}
The parameters $V_{0}$ and $r_{0}$ must be chosen to reproduce the
experimentally observed impurity state at $E_0\approx 110 {\rm meV}$.
In our calculations we have used $V_0= 1.6 eV$ and $r_0 = 2\AA $, but
our results do not depend on this particular choice as long as $r_0$
is sufficiently small. Details of this variational calculation are
presented in  Appendix \ref{append:2Mn}).  In these calculations we
neglected the coupling $G$, since it is much smaller than the binding
energy of the holes.  The final result is that (for $G=0$) the low
lying spectrum of two Mn ions at a separation $R$ can be described by
the following simple effective Hamiltonian,
\begin{equation}
H=-t({R})\sum_\alpha (c_{1\alpha}^{\dagger }c_{2\alpha} +h.c.) + E(R)
\sum_{i=1,2} c_{i\alpha}^{\dagger }c_{i\alpha}\;.
\label{2Mn_eff}
\end{equation}
The parameters $t(R)$ and $\Delta E(R) \equiv E(R)-E_0$ are shown in
Fig.~\ref{comparison_t_fig}. As the inset of the lower panel of
Fig.~\ref{comparison_t_fig} shows, much of the energy shift $\Delta
E(R) $ originates from a simple long-ranged Coulomb shift due to the
Coulomb potential of the neighboring Mn site, $\Delta E_{Coulomb}(R) =
-1/\epsilon R $. The remaining kinetic shift, ${\Delta \tilde E}(R)
\equiv \Delta E(R) + 1/\epsilon R $ is relatively large for small
separations, but vanishes exponentially for large values of $R$.

Having determined the spectrum and the effective Hamiltonian of the
two-Mn ion complex, we can now express the parameters of the effective
Hamiltonian, Eq.~\eqref{eq:H_imp}. The hopping parameter between sites
$i$ and $j$ depends just on their separation, $R_{ij} \equiv |{\bf
  R}_i - {\bf R}_j|$, and is simply $t_{ij}=t(R_{ij})$, the hopping
obtained from the two-Mn problem, while the on-site energies in
Eq.~\eqref{eq:H_imp} are given as \bea E_i&=&E_{0}+ \Delta E_i \;, \nl
\Delta E_i &=& {\sum\limits_{j\neq i}}^\prime \Delta E({R}_{ij})\;.
\label{eq:DeltaE_i}
\eea The sum in this equation should be evaluated carefully: As we
discussed above, for large separations, the shifts $\Delta E(
\mathbf{R}_{i}-\mathbf{R}_{j})$ are dominated by the long-range
Coulomb contributions, and therefore the sum in
Eq.~(\ref{eq:DeltaE_i}) formally diverges. This Coulomb potential is,
however, {\em screened} by the bound valence holes and other charged
impurities in the system, with a screening length of the order of the
typical Mn-Mn distance. Therefore we shall compute the sum in
Eq.~(\ref{eq:DeltaE_i}) with an exponential cut-off, $\Delta
E(R_{ij})\to\Delta E(R_{ij}) \;e^{-R_{ij}/R_{SC}}$, $R_{SC} = 2\;
r_{Mn}$, with ($4\pi r^3_{Mn}/3=n_{Mn}$ the Mn concentration).

Also, our calculation for the hopping matrix elements makes sense only
for nearest neighbor sites. Correspondingly, in the Hopping part of
the Hamiltonian we introduce a rigid cut-off, $R_{0} = 2\; r_{\rm
  Mn}$, to keep only hopping to the first 'shell' of atoms. Our
results do not depend too much on these cut-off parameters; Changing
the Coulomb energy cut-off results in an overall shift of the impurity
and valence band energies, and does not influence the optical
spectrum. Similarly, the results are almost independent of $R_0$ as
long as it is in the range of, $R_0 = 2\; r_{ \rm Mn}$.

\subsubsection{Valence band}

In the tight-binding approach of the previous subsection only the
bound hole states appear. However, to describe optical transitions, we
also need to account for {\em extended states} within the valence
band. These states are involved in local optical transitions, where a
hole localized at site $m$ absorbs a relatively high energy photon
($E\sim200 {meV}$) to make a transition relatively deep into the
valence band.

To account quantitatively for these transitions, we make the following
crucial observations: (i) The transitions involve hole states of
relatively high energy, and correspondingly, states of a relatively
short wavelength. (ii) Since the initial hole states are localized at
the Mn site, the created valence band hole will also be localized
close to it. There the major effect of neighboring Mn sites is to
create a Coulomb potential, which is smooth at the scale of the
wavelength of the created hole.  As a result, we can describe the
final hole state in terms of the following simplified 'semiclassical'
Hamiltonian: 
\be H^{(i)}_{\rm val} \approx -\frac{\gamma }{2}\nabla
^{2} -\frac{1}{\epsilon r_i} + V_{cc}(r_i) +\Delta E_{i,\rm val}\;\;,
\ee where $r_i = |\mathbf{r}-\mathbf{R}_i|$ is the electron coordinate
measured from impurity $i$, and the shift $\Delta E^{i}_{\rm val}$ is
given by 
\be 
\Delta E_{i,\rm val} = - {\sum_j}^\prime
\frac1{\epsilon\;R_{ij}}
\;.  
\ee  
Notice that $\Delta E_{i,\rm val}\approx \Delta E_i$ since  
 most of shift $\Delta E_i$ comes from the Coulomb shift in
Eq.~\eqref{eq:DeltaE_i} (see Fig.~\ref{comparison_delta_e_fig}).
However, these two energies are not exactly  the same,
 $\Delta E_{i,\rm val}\ne \Delta E_i$, since $ \Delta E_i$
also contains the {\em kinetic shift} $\Delta \tilde E$, shown in the inset 
in Fig.~\ref{comparison_delta_e_fig}. It is precisely 
this kinetic  shift, which is responsible for the red-shift 
of the mid-infrared optical peak with increasing Mn concentrations, $x$.

Furthermore, we observe that local optical transitions only involve
$p$-states. These states vanish at the origin, and do not really feel
the central cell correction. Therefore, to describe the local optical
transitions, it is reasonable to include locally these $p$-states
only, and describe the valence band in a second quantized formalism as
\begin{equation}
H_{\rm val}^{\rm band}=
\sum\limits_{m,
i,\alpha }\int\limits_{0}^{\infty }\frac{dk}{\pi }\left( \frac{
  k^{2}}{2m_0} +\Delta E(\mathbf{R}_{i})\right) a_{km\alpha
  ,i}^{\dagger }a_{km\alpha,i}
\label{eq:H_band}\;,
\end{equation}
where $ a_{km\sigma ,i}^{\dagger }$ creates a scattering state in the
$p$-channel around impurity site $i$ in angular momentum channel
$m=x,y,x$, with spin $\alpha$ ($\alpha = -3/2,..,3/2$), and radial
momentum $k$.  The operators $ a_{km\sigma ,i}^{\dagger }$ are
normalized to satisfy the anticommutation relation
\begin{equation}
\{ a_{km\alpha ,i}^{\dagger }, a_{k'm'\beta ,j}\} = \pi
\delta_{\alpha\beta} \delta_{mm'}\delta_{ij} \delta(k-k')\;.
\end{equation}
As mentioned above, scattering states in the $p$-channel do not feel
the central cell correction, therefore the $ a_{km\sigma ,i}^{\dagger
}$'s create to a very good approximation Coulomb scattering states.
Notice that in Eq.~\eqref{eq:H_band} an independent local band is
associated with each impurity site, i.e., propagation between various
impurity sites within the valence band is ignored. This is a
reasonable approximation for  the high-energy, fast processes 
considered in this work.

\subsection{Coupling to the electromagnetic field}

\begin{figure}[t]
\centerline{%
  \includegraphics[width=3.0in,clip]{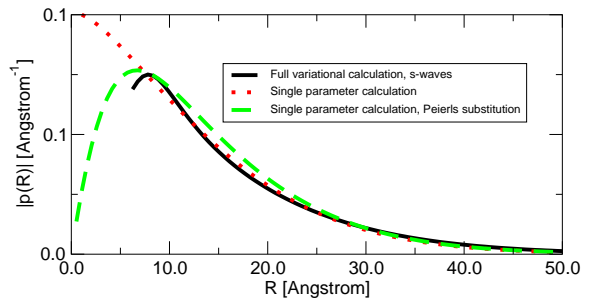}}%
\vspace*{3ex}
\caption{ Momentum matrix element [in units $\hbar = 1$] as a function
  of the separation $R$ between the two Mn sites.  For $R>8\AA$ the
  matrix elements are very well-approximated by the Peierls
  substitution (dashed line).  }
\label{peierls_substitution_single_parameter_full_calculation_fig}
\end{figure}
At the microscopic level, the vector potential couples directly to the
momenta of the valence holes. However, within the impurity band, these
valence holes reside on specific orbitals localized at the Mn sites,
and therefore, as a first step, we need to determine how the vector
potential $\mathbf{A}$ couples to these impurity band states.  To this
end, we determined the matrix elements of the momentum operator
$\mathbf{p}$ between the lowest lying states of our Mn$_2$ molecule
(see Appendix~\ref{append:2Mn}). We find that the effective coupling
of a homogeneous vector potential to the impurity band is given by
\begin{equation}
H_{\rm ext}^{\rm imp} = \sum\limits_{i,j,\alpha } |p(R_{ij})|
{e\gamma\over m_0 c} {\bf A} \cdot {\bf n}_{ij} ( i c_{i\alpha
}^{\dagger }c_{j\alpha } + h.c. ) \;,
\label{eq:H_ext}
\end{equation}
where the unit vector $ {\bf n}_{ij} = \left( {\bf R}_i - {\bf
  R}_j\right) /R_{ij}$ specifies the direction of the bond, and $p(R)$
is the momentum matrix element extracted from the variational
calculation. This matrix element is plotted in
Fig.~\ref{peierls_substitution_single_parameter_full_calculation_fig}.
As we also verified, for $R>8\AA$, this tediously-obtained matrix
element is quite well-approximated by the simple Peierls substitution,
\begin{equation}
p(R) \approx {1\over i}{R \;t(R) \; m_0 / \gamma}\;.
\label{peierls_eq}
\end{equation}
The term \eqref{eq:H_ext} describes {\em intraband transitions}
between states inside the impurity band.  It is this term that
generates the Drude peak and which is responsible for the AC
conductance.

\begin{figure}[t]
\centerline{\includegraphics[width=3.0in,clip]{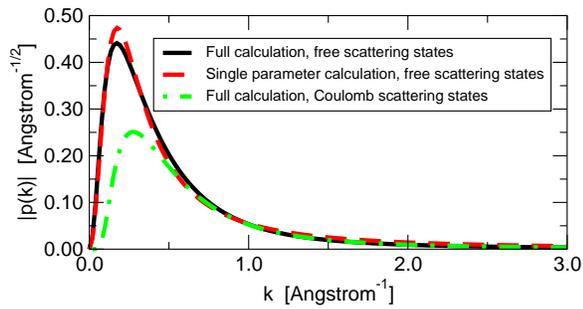}}\vspace*{3ex}
\caption{\label{fig:momentum_ground_state_coulomb} Wave vector
  dependence of the on-site optical matrix element computed for free
  and Coulomb scattering states [$\hbar = 1$]. We also compare this
  with the result obtained for a simple Hydrogen-like variational wave
  function and free scattering states.}
\label{momentum_ground_state_extended_coulomb_fig}
\end{figure}
However, we also need to account for {\em interband transitions},
i.e. transitions from the impurity band to extended states in the
valence band. In our approach, these transitions are local, and are
described by the Hamiltonian,
\begin{equation}
H_{\rm ext}^{\rm tr} = \sum\limits_{i,\alpha,m }
\int\limits_{0}^{\infty }\frac{dk}{\pi } |p(k)| {e\gamma\over m_0 c}
A_m ( i \; a_{k m \alpha ,i}^{\dagger } c_{i\alpha } + h.c. ) \;.
\label{eq:H_ext_tr}
\end{equation}
Here $A_m$ denote the $m=x,y,z$ components of the vector potential
${\bf A}$, and the on-site optical matrix elements $|p(k)|$ can be
computed from the variational wave function of a single Mn site, We
determined $|p(k)|$ for Coulomb scattering states as well as for free
electron states (details of this calculation are presented in Appendix
~\ref{app:1Mn} ). The final results are shown in
Fig.~\ref{momentum_ground_state_extended_coulomb_fig}.

 For free-electron scattering states the single-parameter matrix
 element takes on a particularly simple form,
\begin{equation}
|\langle k|p|\Psi \rangle| =\frac{2^{5/2}}{\sqrt{3}}
\frac{\alpha^{5/2}k^{2}}{(\alpha^{2}+k^{2})^{2}}\;,
\end{equation}
where $\alpha = 0.091\AA^{-1}$. The momentum matrix elements vanish
for $k\to 0$ since $p$-states have a node at the origin, but they also
vanish for very large momenta, where the valence hole wave functions
oscillate much faster than the characteristic scale of the bound
state.  This results in a maximum optical transition rate for valence
states at about $100 \;{\rm meV}$ below the valence band edge. It is
this momentum-dependence of the optical matrix element that is
ultimately responsible for the mid-infrared peak in the optical
spectrum at frequencies $\hbar \omega \approx 200 \;{\rm meV}$.  The
Coulomb and free electron matrix elements behave qualitatively the
same way and are in both cases well approximated by the expression
obtained for a single hydrogen-like variational wave function, $|\Psi
\rangle$. However, the more realistic Coulomb matrix elements have a
somewhat smaller amplitude and the maximum transition rate occurs at
slightly higher values.  In the rest of this paper we shall use these
more realistic Coulomb matrix elements to compute the optical
conductivity.

\section{Optical conductivity at  $T=0$: Mean field approximation}
\label{sec:opt_cond}

In the rest of the paper we shall focus on the calculation of the
optical properties of \gamnas at low temperatures. First, we shall
discuss the case of $T=0$ temperature and treat the Mn spins within
the mean field approximation, justified by the relatively large value
of the Mn spins, $S_{\rm Mn}= 5/2$.  Later in section
\ref{sec:fluctuations} we shall discuss how one can go beyond this
approximation and compute spin wave corrections. These corrections,
however, turn out to be small, and the mean field description turns
out to be quite accurate.  This is related to the fact that, in the
concentration range studied here, the coupling $G$ is much smaller
than the binding energy of the holes $G\ll E_0$ as well as the width
of the impurity band, and therefore it hardly influences the optical
spectrum at these energies.

\subsection{Mean field Hamiltonian}

In our model Hamiltonian, Mn spins couple directly only to states in
the impurity band. Therefore, as a first step, we only need to treat
the coupled Mn spin-impurity band system, described by
Eq.~\eqref{eq:H_imp}.  At the mean field level, the interaction part
is rewritten as \bea G\; {\bf S}_{i} \; c_{i}^{\dagger } {\bf F} c_{i}
&=& G \;\bigl[ \expect{ {\bf S}_{i}} c_i^{\dagger } {\bf F} c_{i} +
  {\bf S}_{i} \expect{ c_{i}^{\dagger } {\bf F} c_{i} } \nl &-&
  \expect{ {\bf S}_{i}} \expect{ c_{i}^{\dagger } {\bf F} c_{i}
  }\bigr] +H_{\rm fluct}(i) ;, \\ H_{\rm fluct}(i) &=& G \; ({\bf
  S}_{i}-\expect{ {\bf S}_{i}}) ( { c_{i}^{\dagger } {\bf F} c_{i} } -
\expect{ c_{i}^{\dagger } {\bf F} c_{i} } )\;,
\label{eq:fluct}
%
\eea and the fluctuating part, $ H_{\rm fluct}$ is neglected. With
this approximation, Eq.~\eqref{eq:H_imp} becomes exactly solvable at
any temperature.  For given values of $\expect{ {\bf S}_{i}}$, the
hole part of the Hamiltonian is diagonalized by the unitary
transformation
\begin{equation}
a_{\mu}^{\dagger} = \sum_{i, \alpha}\phi_{\mu}(i \alpha)\,
c_{i\alpha}^{\dagger },\; \;\;\;\;\; c_{i\alpha}^{\dagger} =
\sum_{\mu}\phi_{\mu}^*(i \alpha)\, a_{n}^{\dagger
}\;,\label{eq:unitary_transformations}
\end{equation}
where wave functions $\phi_{n}^*(i \alpha) $ can be found by solving a
relatively simple eigenvalue equation. The mean field Hamiltonian can
then be rewritten in terms of the operators $a_{\mu}^{\dagger}$ and
the corresponding eigenenergies, $E_\mu$, as \be H^{\rm MF}_{\rm imp}
= \sum_\mu E_\mu a_{\mu}^{\dagger}a_{\mu} - \sum_i {\bf S}_{i} {\bf
  h}_i + \sum_i \expect{ {\bf S}_{i}} {\bf h}_i \;,
\label{eq:H_MF}
\ee where the local fields ${\bf h}_i = - G \expect{ c_{i}^{\dagger }
  {\bf F} c_{i}} $ can be expressed in terms of the new basis as \bea
    {\bf h}_i &=& - G\sum_{\mu } {\bf F}_{\mu \mu}(i) \; f(E_\mu)\;,
    \\ {\bf F}_{\mu\nu}(i) &\equiv& \sum_{\alpha \beta}
    \phi_{\mu}^{*}(i \alpha) {\bf F}_{\alpha \beta}\,\phi_{\nu}(i
    \beta)\;,
\label{eq:F_munu}
\eea with $f$ the Fermi function. Trivially, ${\bf h}_i$ determines
the expectation value $\expect{ {\bf S}_{i}}$, that enters the
eigenvalue equation of $\phi_{\nu}(i \alpha)$. This field must
therefore be determined selfconsistently. At $T=0$ temperature the
selfconsistency equations become rather simple, since then the spins
are completely polarized in the direction ${\bf h}_i$, $\expect{{\bf
    S}_{i}} = S_{\rm Mn} {\bf h}_i/|{\bf h}_i|$, and the Fermi
function simply becomes a step function. However, for small
concentrations  and/or finite temperature the Mn spins are not fully
polarized, and the numerical 
solution of the mean field equations becomes time-consuming.

\subsection{Optical conductivity}

The mean field solution of the Hamiltonian provides us the equilibrium
state of the coupled spin-impurity band system. To compute the optical
spectrum, however, we also need to transform the coupling to the
electromagnetic field to the 'canonical' basis,
Eq.~\eqref{eq:unitary_transformations}.
 
We can express the coupling of the impurity band to a time-dependent
vector potential ${\bf A}(t)$ in terms of the canonical operators,
$a_\mu$ as $ H_{\rm ext}^{\rm imp} = \sum_{\mu,\nu} {\bf A}\cdot
a^\dagger_\mu {\bf j}_{\mu\nu}a_\nu $, where the current operator's
matrix element is defined as \bea {\bf j}_{ \mu\nu} =
\sum_{i,j,\alpha}\left | p(R_{ij})\right |\frac{e \gamma}{mc}\cdot
    {\bf n}_{ij}\times \nl \left[ i\; \phi_{\nu}^*(i
      \alpha)\phi_{\mu}(j\alpha) - i \;\phi_{\nu}^*(j
      \alpha)\phi_{\mu}(i\alpha) \right]\;.
\label{eq:j_matrix_imp}
\eea Similarly, we rewrite the term \eqref{eq:H_ext_tr} in the
Hamiltonian in this canonical basis as
\begin{equation}
H_{\rm ext}^{\rm tr} = \sum_{km\alpha,i,\nu}A_{m}(t)\left( {\cal
  J}_{k, \alpha i,\nu}a^{\dagger}_{km\alpha,i}a_\nu +{\rm h.c.}\right
)\;,
\label{eq:j_matrix_impval}
\end{equation}
with ${\cal J}_{k,\alpha i, n} = \left | p(k) \right |{e \gamma \over
  mc} i\phi_n(i \alpha)$.

The optical conductivity $\sigma(\omega)$ can be computed by means of
the Kubo formula, which expresses it in terms of the retarded
current-current correlation function,
\begin{equation}
\sigma\left(\omega\right) =- \frac{1}{\omega} {\rm Im}\;\Pi_{JJ}^{\rm
  ret}(\omega)\;\label{eq:current_correlator}.
\end{equation}
Since, corresponding to Eqs.~\eqref{eq:j_matrix_imp} and
Eq.~\eqref{eq:j_matrix_impval}, our current operator consists of an
interband and an intraband part, the current-current correlation
function can also be expressed in terms of two contributions: an {\em
  intraband contribution}, $\sigma_{\rm intra}$, corresponding to
transitions within the impurity band, and an {\em interband
  contribution}, $\sigma_{\rm inter}$, describing transitions from the
impurity band to the valence band.  In our approach the low frequency
behavior arises from impurity band scattering alone, while the valence
band plays practically no role there. The high frequency conductivity,
on the other hand, is typically dominated by interband transitions to
the valence band. In the present, impurity band approach, these
transitions give rise to a mid-infrared peak close to $2000 cm^{-1}$.

\begin{widetext}
The previously mentioned two contributions can be easily computed
within the mean field approach, using the diagrammatic formalism
presented in Section~\ref{sec:fluctuations}, and the interband
contribution can be simply expressed as \be \sigma_{\rm
  inter}\left(\omega\right) = -\frac{1}{\omega} \sum_{k,
  \alpha}\sum_{j, \mu} \left | {\cal J}_{k,\alpha i, \mu} \right |^2
\;{\rm Im}\;\left\{ \frac{f(E_{\mu})-
  f(\varepsilon_k(i))}{\omega-\varepsilon_k(i)+E_{\mu}+i\delta}\right\}\;.
\label{eq:inter}
\ee Notice that the transition being local, it is also the local
energy of a valence band hole, that appears in the denominator of this
expression, $\varepsilon_k(i) = \frac{ k^{2}}{2m_0} +\Delta E_i$.

The intraband contribution can be written in a similar way as \be
\sigma_{intra}\left(\omega\right) = - \frac{1}{\omega} \sum_{\mu, \nu}
|{\bf j}_{\nu \mu}{\bf e}_0|^2
\; {\rm Im}\;\left\{ \frac{f(E_\mu)-
  f(E_\nu)}{\omega+E_\mu-E_\nu+i\delta}
\right\}\;,\label{eq:intra} \ee with ${\bf e}_0$ the polarization of
the external light.
\end{widetext}

Clearly, both the energy and the precise structure (i.e., localized or
extended character) of the impurity band states enter the matrix
elements in the above expressions. However, apart from the value of
the DC conductance, which clearly must vanish as one approaches the
localization transition, the gross high-frequency features will turn
out to  not be  very sensitive to the localization transition
itself. This is not very surprising, since the localization transition
involves mostly states at the Fermi energy, while optical transitions
are dominated by states deep below or high above the Fermi energy.

\section{Mean field results}
\label{sec:mean_field_res}

\subsection{Density of states}

\begin{figure}[b]
\centering
\includegraphics[width=3.0in,clip]{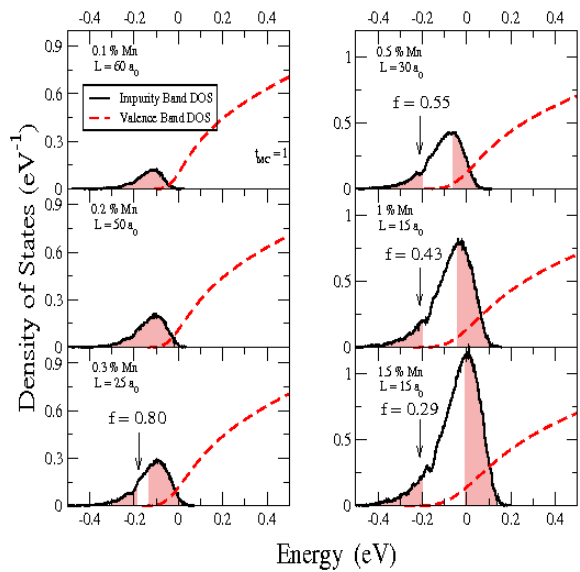}
\caption{ Ground state density of states for different ${\rm Mn}$
  concentrations and a fixed hole fraction, $f=0.5$.  The solid lines
  represent the DOS for the impurity band while dotted lines denote
  the valence band DOS. The Monte Carlo relaxation time is fixed to
  $t_{MC}= 1$ per Mn ion. Shaded areas indicate localized states,
  while unshaded regions correspond to extended states.  For these
  concentrations we find two mobility edges in the impurity band: one
  separating states in the tail of the impurity band, while the other
  separating localized states in the impurity band - valence band
  gap.}
\label{fig:dos_mobility_edge}
\end{figure}

\begin{figure}[b]
\centering
\includegraphics[width=3.0in,clip]{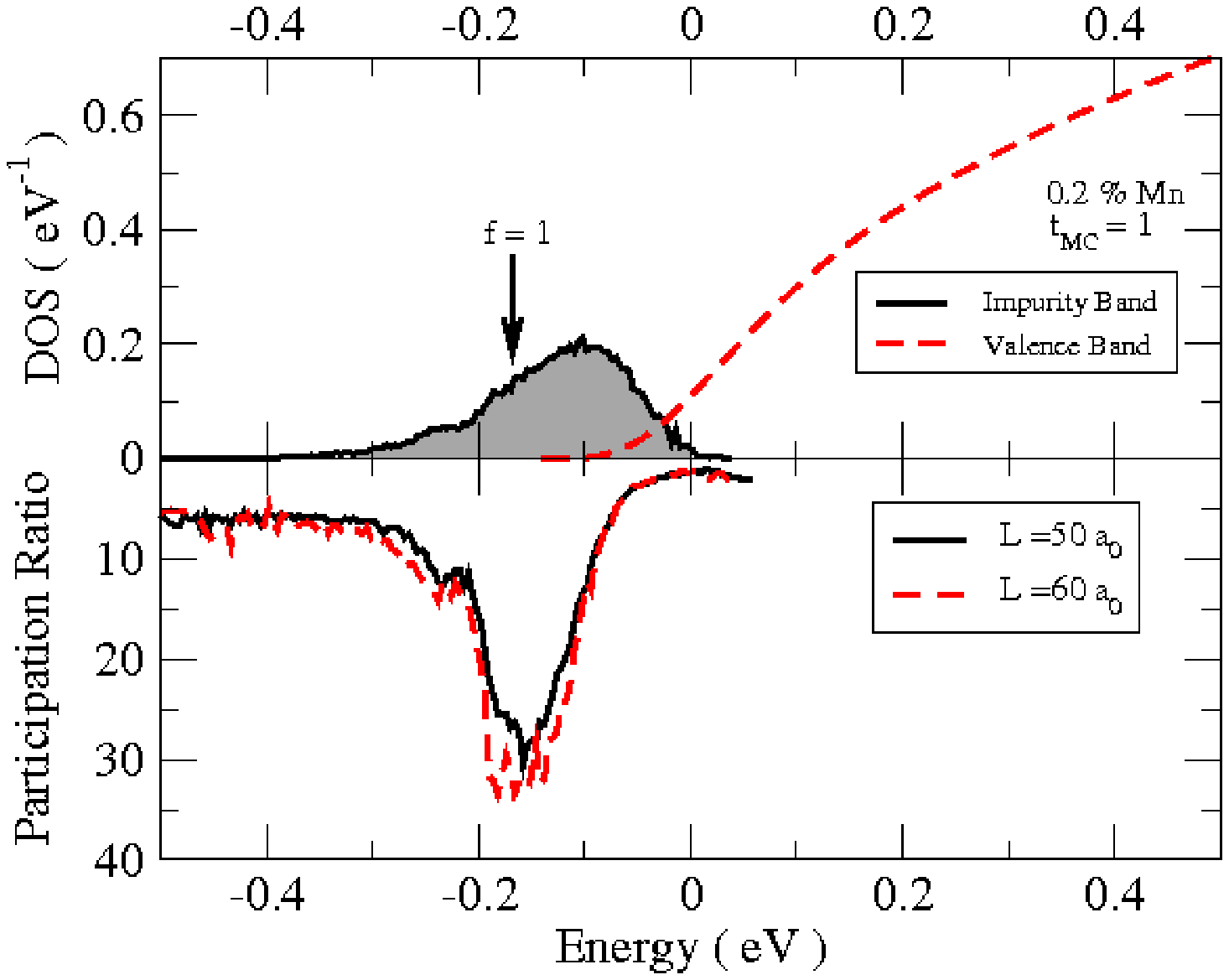}
\includegraphics[width=3.0in,clip]{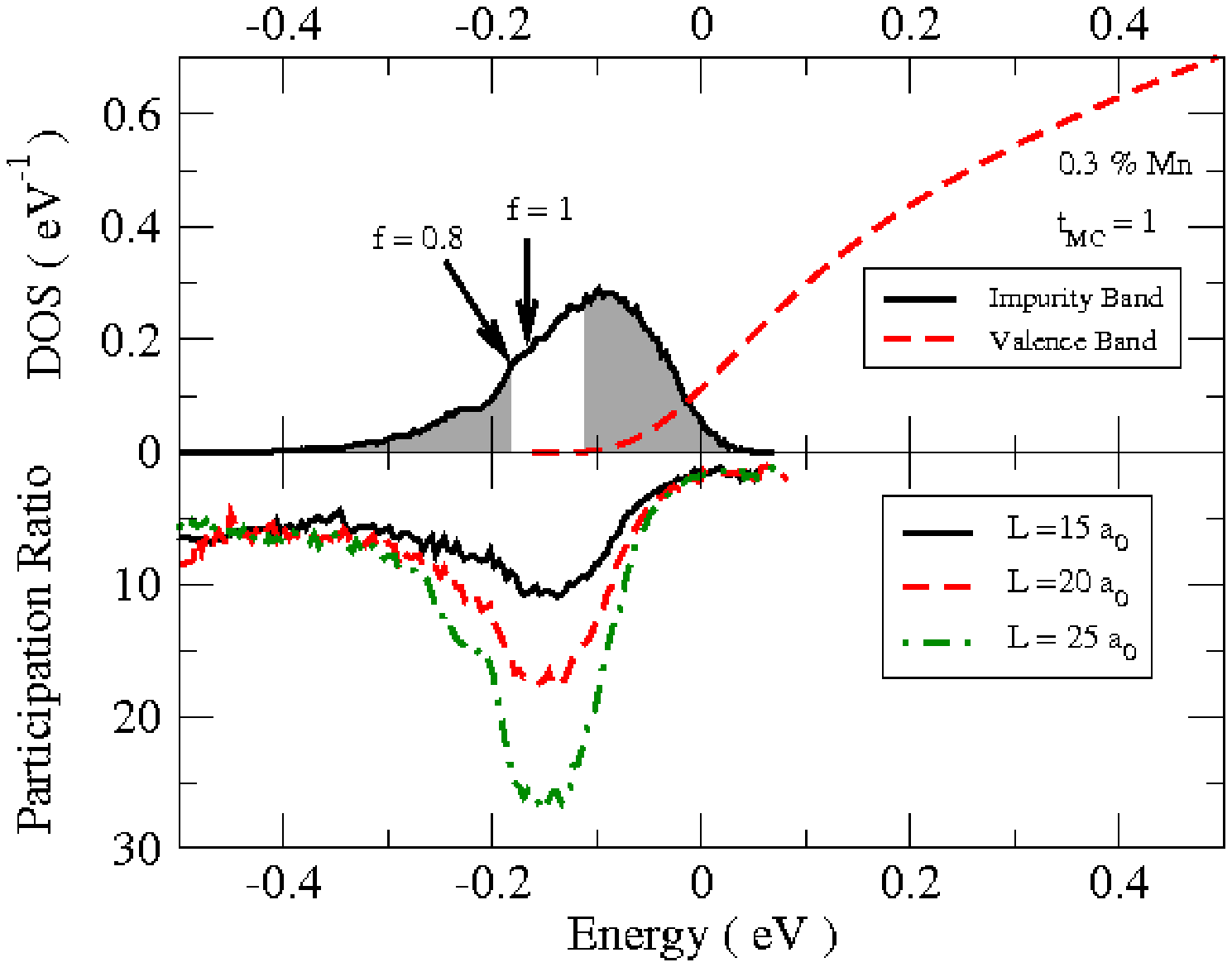}
\caption{ Ground state density of states and participation ratios for
  different ${\rm Mn}$ concentrations but a fixed hole fraction,
  $f=0.5$.  The Monte Carlo relaxation time is fixed to $t_{MC}=
  1$. The shaded regions indicate the position of the mobility edge
  for each concentration.}
\label{fig:dos_pr}
\end{figure}

Let us start first by discussing some details of the numerical
solution of the mean field equations described in the previous
section, and the structure of the hole states obtained.  To start the
numerical calculation, we first generate an initial distribution for
the substitutional ${\rm Mn}$ impurities on an fcc lattice with
lattice constant $a\approx 5.6\;\AA$. Nearest neighbor Mn sites are,
however, not favored during the growth process, since Mn ions act as
charged impurities. To simulate this effect, we let the Mn ions relax
through a simple classical Monte Carlo diffusion process while
assuming a screened Coulomb interactions between them.\cite{Fiete}
Typical values of this Monte Carlo time used are in the range of
$t_{\rm MC}\sim 1 \;{\rm step / atom}$.  In all our calculations we
use periodic boundary conditions to suppress surface effects.

Once the configuration of the ions is determined, we can construct the
effective Hamiltonians Eq.~\eqref{eq:H_imp} and \eqref{eq:H_band} as
described in Section~\ref{sub:EffectiveHamiltonian}, Having
constructed the Hamiltonian, we then solve the mean field equations
iteratively and determine the equilibrium spin configuration.  Here we
only focus on $T=0$ temperature, where $\expect{ {\bf S}_{i}} = S {\bf
  \Omega}_i$, with ${\bf \Omega}_i$ acting simply as a classical
variable. In the end of the mean field self-consistency loop the
states in the impurity and valence band are fully characterized, i.e.
the corresponding eigen-energies $E_\mu$  and
$\varepsilon_k(i)$ as well as the wave functions are available. We can
then compute the necessary matrix elements of the current operator
using Eq.~\eqref{eq:H_ext}, and apply \eqref{eq:current_correlator}
and \eqref{eq:intraband} immediately to compute the contributions to
the optical conductivity.

For a good accuracy, we consider system sizes $L$ as large as
possible, and we average over many configurations. Usually for a given
${\rm Mn}$ concentration and hole fraction, we average over 100
different impurity configurations, and we consider systems with the
number of impurities in the range of 200.

In Fig. \ref{fig:dos_mobility_edge} we present results for the
impurity band density of states (DOS) for different ${\rm Mn}$
concentrations but for a fixed hole fraction, $f=0.5$.  We also
indicate the average DOS for the valence band.  Shaded regions,
denoting localized states were determined by by analyzing the finite
size scaling of the participation ratio
\begin{equation}
{\rm PR}(\mu) = \left [ \sum_i \left (\sum_{\alpha} \mid \phi_{\mu}(i
  \alpha) \mid^2 \right) \right]^{-1}\;.\label{eq:pr}
\end{equation}
For extended states ${\rm PR}$ scales with the system size, while for
the localized states remains finite for $L\to
\infty$. Fig. \ref{fig:dos_pr} shows the details of such a finite size
scaling analysis.

As we expect, increasing the ${\rm Mn}$ concentration, the impurity
band gets broader and starts to overlap with the valence band for
${\rm Mn}$ concentration as low as 0.5\% (see also
Fig.\ref{fig:dos_mobility_edge}). However, at the same time increasing
Coulomb disorder shifts the tail of the impurity band deeper inside
the gap (the top of the valence band is fixed at $E_V^{top}=0$).
States inside this tail are strongly localized and have an
overwhelming localized impurity state character.

We clearly observe a metal insulator transition (MIT) at a critical
concentration $x \approx 0.2$\%, which corresponds to a hole
concentration $p\simeq 4.5\times 10^{19} {\rm cm^{-3}}$ for $f\approx
1$.  In our approach this metal-insulator transition is simply an
Anderson localization transition, since electron-electron interaction
effects are neglected: For $ x< 0.2$\% all the states in the impurity
band are localized by disorder and the system is an insulator, while
for $x > 0.2$\% the metallicity of the system still depends on the
position of the Fermi level inside the impurity band. As an example
(see Fig.\ref{fig:dos_pr}, bottom layer) for $ x = 0.3\%$ and hole
fractions less than $f=0.8$ the system is still an insulator, while
for a hole fraction $f\sim 1.0$, the Fermi level is already inside the
delocalized region and we find a metallic state.  Increasing the ${\rm
  Mn}$ concentration to larger values, the extended region grows and
the system has a metallic behavior for typical values of the hole
fraction, $f\sim 0.5\div 1$, (see also
Fig. \ref{fig:dos_mobility_edge}).  For hole fractions $f\approx 0.6$,
e.g., we find that the ground state is metallic for $x>0.5\%$.  These
numbers are in very good agreement with experimental results, where
the MIT has been reported to occur in a concentration range $x\approx
0.2-0.3$\%.\cite{Jungwirth3,Moriya}

We must remark that here we completely neglected electron-electron
interaction.  Electron-electron interaction can obviously modify many
of the physical properties close to the MIT and lead to the appearance
of a Coulomb gap as well as Altshuler-Aronov-type
anomalies.\cite{Altshuler} On the other hand, even annealed \gamnas
samples are typically compensated, and the electron-electron
interaction has been found to be less relevant for such systems.
Also, we can argue that our results are in a way self-consistent, and
probably not very sensitive to Coulomb interaction;
For typical concentrations $x<3 \%$ and hole fractions in the range $f\simeq
0.3\div 1$, we find that 
aproximately $ 60\div 80\;\%$ of the sites have less then one electrons on them, 
and less then $\sim 5\; \%$ of sites are more then doubly occupied.
 Furthermore, even for the most localized orbitals, the
participation ratio is rather large $\sim 6-10$.  Therefore, a typical
site is occupied by a single electron and the Coulomb correlations are
not too relevant.\cite{Fiete} Treating the Coulomb interaction
appropriately is, however, a real theoretical challenge even for much
simpler model systems, and is beyond the scope of the present work.

\begin{figure}[t]
\centering \includegraphics[width=3.0in,clip]{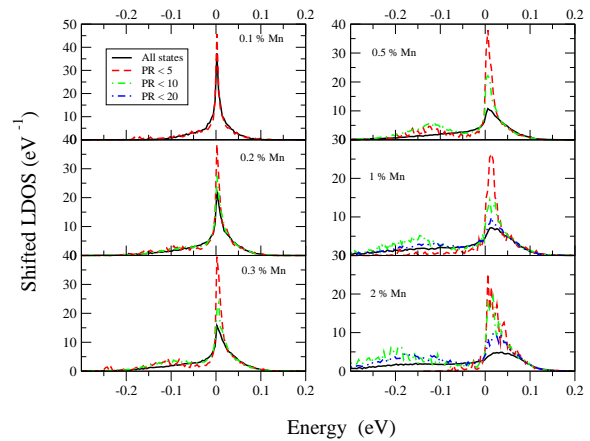}
\caption{Shifted Local density of states for different ${\rm Mn}$
  concentrations.  The Monte Carlo relaxation time is fixed to
  $t_{MC}= 1$.}
\label{fig:shifted_dos}
\end{figure}

 Let us close this subsection by presenting a rather curious quantity
 that we termed 'shifted local density of states' (shifted-LDOS),
 where at every site we computed the local density of states by
 subtracting the binding energy of a hole, $E_0$, as well as the value
 of the average Coulomb shift. While the unshifted LDOS is rather
 featureless, this quantity displays a sharp peak around zero energy.
 A more detailed analysis reveals that this sharp peak is associated
 with localized states having a small participation ratio, i.e. states
 in the tail of the impurity band. To show this, we also plotted the
 contribution of states with ${\rm PR}$'s smaller than a given value
 to the shifted-DOS. Clearly, the peak observed at zero energy is
 entirely due to localized states deep inside the gap.  The physical
 interpretation of this peak is thus plausible; States in the tail of
 the impurity band are generated by random and large fluctuations of
 the Coulomb disorder.  These states are simply localized impurity
 states deep inside the gap that are being mostly occupied, and they
 give rise to local optical transitions which have a strong impurity
 state transition character.
%

\subsection{Optical conductivity}

\begin{figure}[t]
\centering
\includegraphics[width=3.3in,clip]{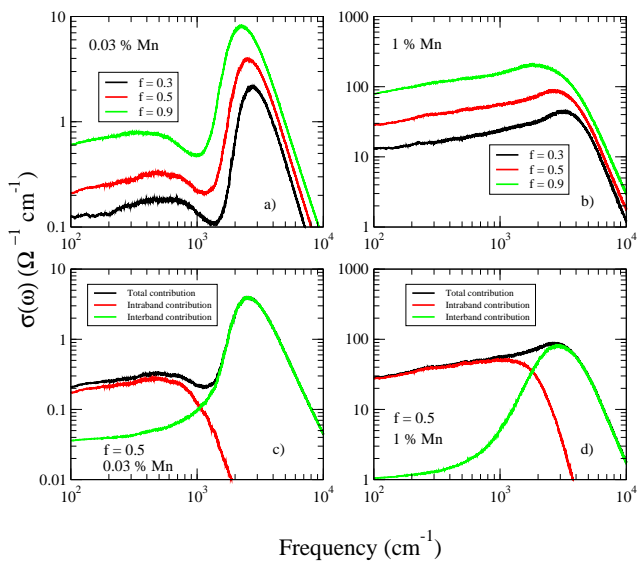}
\caption{The real part of the optical conductivity for different ${\rm
    Mn}$ concentrations and hole fractions. In a) and c) the ${\rm
    Mn}$ concentrations is $0.03\%$, while in b) and d) it is $x=1\%$.
  In c) and d) the intraband and interband contributions are presented
  for a hole fraction $f=0.5$.  The sample in  panel a)
  is insulating 
  while the one in  panel b) is metallic.  The peak
  observed at $\sim 
  2000\div 2500 {\rm cm^{-1}}$ for the metallic sample is due to the
  valence to impurity band transitions while the broad feature at
  smaller energies is a reminiscent Drude peak generated by the
  impurity band contributions.}
\label{fig:optical_conductivity}
\end{figure}


Having diagonalized the mean field Hamiltonian, it is straightforward
to compute the optical conductivity. As we shall see, our theory,
which incorporates inter-impurity band transitions as well as
intraband transitions, accounts qualitatively as well as
quantitatively for all major features of the experimentally observed
optical spectrum. We shall also present a scaling analysis of the
optical conductivity close to the metal-insulator transition and
compute the dynamical critical exponent.

The Drude contribution, i.e. the $\omega\to0$ feature is controlled by
excitations close to the Fermi level.  Since in this small
concentration limit the Fermi level resides in the impurity band, the
``Drude peak'' and also the DC conductivity are expected to be
dominated by the {\em intra impurity band contribution}.  The
contributions of extended valence band states is negligible for
$\omega\ll E_0$, since these states are too far from the Fermi level
to be of any relevance for the Drude contribution.

The interband contribution, on the other hand, is generated by
transitions between the impurity and valence bands.  As can be seen in
Fig. \ref{fig:MIT}, the two bands start to overlap for ${\rm Mn}$
concentrations $x>0.5 \%$. However, although it also depends on the
hole fraction $f$, the energy gap between the Fermi level and the top
of the valence band is always in the range $\Delta = 0.1 eV$. We
therefore expect to see a broad mid-infrared feature in the energy
range $\Delta <\omega <2 \Delta$.

These expectations are indeed met by our numerical results shown in
Fig. \ref{fig:optical_conductivity}.  In panel d) we also display the
intraband and interband contributions separately for a typical ${\rm
  Ga_{1-x}Mn_{x}As }$ sample in the metallic regime, with $x= 1 \%$
and a hole fraction $f=0.5$.  The intraband contribution has a
Drude-like behavior, while the interband signal presents a peak at the
energy $2500 {\rm cm^{-1}}$.  However, the Drude contribution is
rather broad, and in fact, already for this relatively small
concentration, it almost completely merges with the interband
contribution.  The overall behavior is in good agreement with the
experimental data presented in Ref.[\onlinecite{Singley1, Singley2,
    Burch1, Burch2}]. Remarkably, not only the peak position but also
the overall magnitude work out to have values in the experimental
range.

As already shown in Fig.~\ref{fig:peak_positions}, increasing the
number of carriers leads to a red-shift of the mid-infrared peak in
agreement with the experimental observations; more metallic samples
tend to have a strongly red-shifted mid-infrared peak at a frequency,
$\sim 1500 \;{\rm cm}^{-1}$ rather than at $\sim 2500\;{\rm cm}^{-1}$.
We find that in the concentration range considered here, to a good
approximation, the red shift only depends on the effective optical
spectral weight, $N_{\rm eff}$, defined through Eq.~\eqref{eq:N_eff}.
Remarkably, there is an almost perfect match between our theoretical
results and the experimental values for intermediate concentrations.
The decreasing trend observed in the position of the mid-infrared peak
as function of hole fraction can be understood simply from
Fig. \ref{fig:peak_positions}: for a given ${\rm Mn}$ concentration,
increasing the carrier numbers shifts the Fermi level closer to the
top of the valence band. As a consequence, the optical gap decreases
and the mid-infrared peak moves towards lower energies.

A first look at the results presented in panels c) and d) of Figure
\ref{fig:optical_conductivity} can drive us to the conclusion that
there is not much difference between  the overall optical response
of the metallic and insulating samples. However, the effective mass
analysis and spectral weight analysis presented in the following
subsection clearly shows the difference between these phases.

However, before turning to the effective mass analysis, let us discuss
the low frequency properties of the optical response close to the MIT
transition.  The scaling theory of Abrahams  {\it et al.}
\cite{Abrahams} has been extended to the dynamical conductivity by
Shapiro {\it et al.},\cite{Shapiro} who showed that at the mobility
edge the ac conductivity of a $d=3$-dimensional system obeys a power
law, $\sigma(\omega) \sim \omega^{d-2/d}$. They also found that on the
metallic side the conductivity goes as $\sigma(\omega)-\sigma(0) \sim
\omega^{(d-2)/2} $ while in the insulating phase $\sigma(\omega) \sim
\omega^{2}$.

\begin{figure}[t]
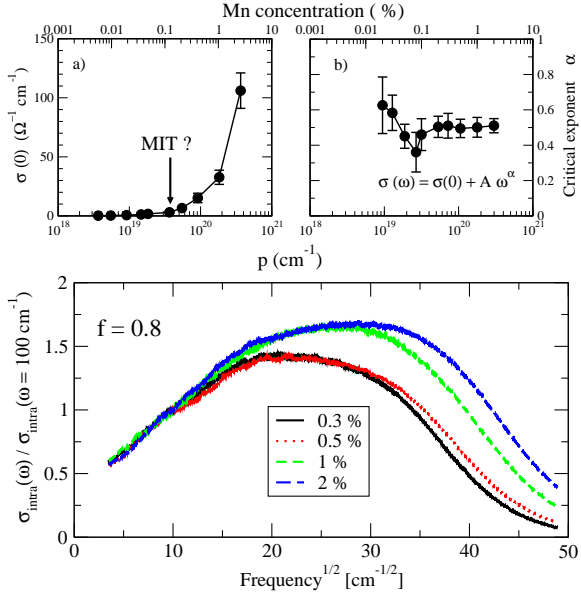

\centering \includegraphics[width=3.0in,clip]{scaling.eps}
\includegraphics[width=3.0in,clip]{optical_conductivity_linear_linear.eps}
\caption{ Upper panels: a) the dc-limit of the intraband optical
  conductivity for concentrations between $0.02 \%$ and $1 \%$.  b)
  The scaling exponent $\alpha$ for the conductivity in the low energy
  limit as function of the carrier concentration. The MIT transition
  is indicated by a drop of the critical coefficient $\alpha$ down to
  0.35.   Bottom panel: Frequency dependence of the
  intraband optical 
  conductivity in the small frequency limit presented for different
  ${\rm Mn }$ concentrations.  In all three figures the hole fraction
  was fixed to f = 0.8.}
\label{fig:scaling}
\end{figure}
 To our knowledge, the first numerical evaluation of the optical
 conductivity close to the localization transition based on the Kubo
 formula was done in Ref. [\onlinecite{Lambrianides}], where indeed it
 was shown that the dynamical exponent for the optical conductivity of
 a three-dimensional Anderson model at the critical point is
 $1/3$. More recently, the analysis was extended to unitary and
 symplectic systems \cite{Shima} where the same exponent was found.
 In the lower panel of Fig.~\ref{fig:scaling} we present the low
 frequency limit of the intraband optical conductivity close to the
 MIT transition for a fixed hole fraction $f=0.8$, where the critical
 carrier concentration is found to be $p_c\simeq 4\times 10^{19} {\rm
   cm^{-3}}$ ($x=0.2$) by a participation ratio analysis. For $p>p_C$
 the low energy optical data can be nicely described by a power law,
 $\sigma (\omega) = \sigma(0)+ A\omega^{\alpha}$.  Since we use
 periodic boundary conditions, we observe a continuous metal-insulator
 transition similar to the case of the Anderson model.\cite{Weisse}
 For an infinite system, the dc conductivity $\sigma (0)$ approaches
 zero at the transition point and vanishes in the insulating phase.
 Of course, for a system of a finite size like ours, $\sigma (0)$ is
 not strictly zero even in the insulating phase, and only a crossover
 is observed between these two regimes (see the upper left panel of
 Fig.~\ref{fig:scaling}).

The most relevant quantity is the exponent $\alpha$. In the metallic
state a best fit gives $\alpha \sim 0.5$ in good agreement with the
analytical predictions. Although the error bars are rather large,
$\alpha$ seems to slightly decrease down to $0.35$ at the MIT where it
starts to increase again in the insulating regime. The two approaches,
the participation ratio and the scaling analysis of the dynamical
conductivity based on which we have identified the MIT point are in
good agreement. In both cases the critical point is associated with
approximately the same hole carrier concentration.

\subsection{Effective mass analysis}

\begin{figure}[b]
\centering \includegraphics[width=3.5in,clip]{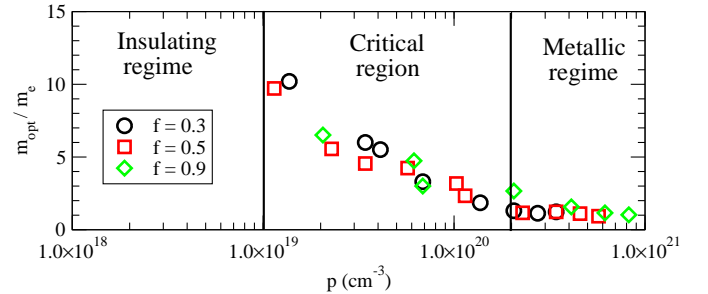}
\caption{ The optical mass as function of number of carries. Only the
  impurity band contribution was used to compute $\mopt$.  Deep in the
  metallic regime, for carrier concentrations larger than $2 \times
  10^{20} {\rm cm^{-3}}$ the effective mass remains practically
  constant, while in the critical regime as approaching the MIT
  transition ($p\simeq 10^{19}\div 10^{20} {\rm cm^{3}}$) an
  power-like increase in the effective mass is observed.  }
\label{fig:effective_mass}
\end{figure}

Experimentally, optical conductivity data are often used to extract
important information on the charge carriers such as their
concentration or effective mass, or the value of $k_Fl$.  From the
analysis of the experimental data, two types of effective masses can
be extracted: $(i)$ the effective mass $\mstar$ that enters the Drude
formula for the resistivity, i.e.  the dc limit of the optical
conductivity.  $(ii)$ the optical mass $\mopt$ that is related to the
spectral sum rule.  These two masses are often quite different.  Also,
they are typically extracted based upon some assumptions (isotropical
mass and quadratic Hamiltonian, etc.), which are violated in the real
system.  Nevertheless, they both contain important information, and it
is therefore interesting to determine them from our theoretical
results and compare to the experimental data.

The effective mass $\mstar$ can only be roughly estimated from 
the data. Following the procedure of Ref~[\onlinecite{Burch1}], 
we typically find it to be in the rage of $\mstar\sim 10-30\;m_e$. 
The mass $\mstar$ is also  related to the the metallicity
parameter, $k_F l$, with $k_F$ the Fermi momentum and $l$ the mean
free path. In fact, this dimensionless parameter can be extracted much
more reliably from the experimental data than $\mstar$, and it has a
much more obvious interpretation: Small values of $k_F l$ correspond to
samples in the insulating phase or close to the metal insulator
transition, while good metals are characterized by large values of
$k_Fl$. We shall therefore focus on this parameter.

The value of $k_Fl$ can be estimated from the hole concentration and
the dc conductivity using the Drude formula, rewritten as \be k_F l =
\left (\frac{h}{e^2}\right )\sigma(0) \frac{3}{2s}\lambda_F\;.  \ee
Here $s$ is the spin degeneracy of the band. In an $s=4$ four band
model we found that for carrier concentrations between $p =2.0 \times
10^{20}\div 9.0 \times 10^{20} {\rm cm^{-1}} $ our optical
conductivity would correspond to values $0.3< k_F l < 3.0$. In other
words, for these high resistances the mean free path is already less
than the Fermi wavelength $\lambda_F = 2\pi / k_F$. Thus the Drude
analysis clearly leads to a inconsistency, and shows that the optical
conductivity does not originate from a weakly perturbed valence
band. While in the present work, where we use an impurity band
approach, this is obvious, the same analysis can also be performed for
the experimental data to show, that a weakly perturbed valence band
picture is inappropriate to describe most \gamnas
samples.\cite{Zarand} The dc conductivity values obtained within
our impurity band picture are, on the other hand, perfectly consistent
with the experimentally extracted values of $k_Fl$.

The optical mass, $\mopt$, can be defined through the effective optical
spectral weight, $N_{\rm eff}$, defined by the integral
\begin{equation}
{p \over \mopt} \equiv N_{\rm eff}\equiv \frac{2}{\pi e^2} 
\int_0^{\omega_c}
\sigma_{\rm intra}\left(\omega \right) d\omega\;,
\label{eq:N_eff}
\end{equation}
with $\omega_c$ a somewhat arbitrary energy cut-off, which has 
been set to $800\; {\rm meV}$ in Ref.~[\onlinecite{Burch1}].
 The optical mass $\mopt$ defined through this formula can be quite
 different from the microscopic mass of the carriers, $m_0$,
and it provides a useful measure of the ``heavyness'' of
 the charge carriers.  To compare with the experiments, we extracted
 $\mopt$ from our data, and in Fig. \ref{fig:effective_mass} we
 plotted it as a function of carrier concentration.  
Deep in the
 metallic regime, we have obtained an optical mass that remains
 approximately constant, $ \mopt /m_e \approx 1 $, for carrier
 concentrations larger than $p\approx 2\times 10^{20}$, i.e. for ${\rm
   Mn}$ concentrations larger than about $x \approx 1\% $. Our $k_Fl$
 values as well as our optical mass results are in good agreement with
 the experimental data,\cite{Burch1}
where the optical mass was found to be in the range
$0.7<\mopt/m_e<1.4$. Notice that these values are about twice as large
as the bare valence band mass, $m_0 = 0.56\; m_e$.  As we approach the
MIT transition, $\mopt$ increases rapidly, and can reach values as large
as $\mopt \approx 10\; m_e$. This large optical mass renormalization is
characteristic of the vicinity of the critical concentration, (see
Fig.\ref{fig:effective_mass}).  In this regime we find a metallicity
parameter $k_F l$ that is always less than $0.5$.

\section{Spin fluctuations}
\label{sec:fluctuations}

So far we have treated Mn spins at the mean field level. In the
present section we shall discuss how one can go beyond mean field by
systematically including spin fluctuations. To do that, we shall make
a $1/S$ expansion around the classical limit, and represent spin
fluctuations using Holstein-Primakoff bosons.\cite{Konig} The ground
state is, however, generally
non-collinear,\cite{Schliemann,ZarandJanko} leading to some
computational  complications.
 
Here we shall consider again only the $T=0$ temperature limit. There
the mean field equations imply that $\expect{{\bf S}_{i}} = S_{\rm
  Mn}{\bf e}_i^z$, with the unit vector ${\bf e}_i^z = {\bf h}_i/|{\bf
  h}_i|$ being parallel to the external field created by the valence
holes. We shall use this direction as a quantization axis of the Mn
spin at site $i$, and introduce two additional unit vectors, ${\bf
  e}_i^{x,y}$ perpendicular to ${\bf e}_i^z$.  Using
Holstein-Primakoff bosons, we can then represent the Mn spin operators
at site $i$ as follows,
\begin{eqnarray}
{\bf e}_i^z \cdot {\bf S}(i) & = & S - b^{\dagger}_{i}b_{i} \nl {\bf
  e}_i^x \cdot {\bf S}(i) & = & \sqrt{\frac {S}2}\,\, ( b_{i} +
b^{\dagger}_{i}) +\dots\;, \nl {\bf e}_i^y \cdot {\bf S}(i) & = & i \;
\sqrt{\frac {S}2}\,\, ( b^{\dagger}_{i} - b_{i}) + \dots \;,
\end{eqnarray}
where we have neglected terms that are subleading in $1/S$.

\begin{table}[t]
\begin{tabular}{lrl}
\hline \hline
\includegraphics[width=1in]{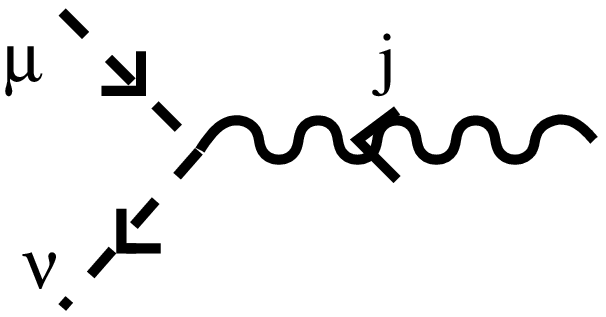} &
$\gamma^{+}_{\mu, \nu}\left (j\right ) $\vspace{0.3in} \null&
\\ \includegraphics[width=1in]{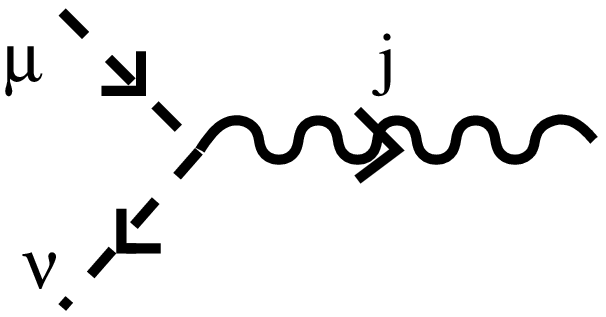}
&$\gamma^{-}_{\mu, \nu}\left (j\right )$ \vspace{0.3in}&
\\ \includegraphics[width=1in]{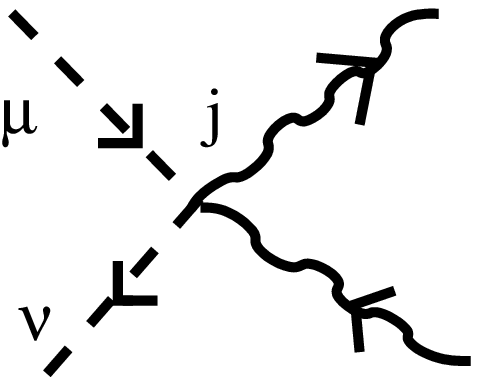}
&$\gamma^{z}_{\mu, \nu}(j)$ \vspace{0.3in}& \\
\hline \hline
\end{tabular}
\caption{Vertices functions within the interacting field theory.
  Dashed lines represent impurity hole propagators, while wavy lines
  denote bosonic propagators. }
\label{table:green_functions}
\end{table}
In this language, the mean field Hamiltonian, Eq.~\eqref{eq:H_MF}
simply becomes \be H^{\rm MF}_{\rm imp} = \sum_\mu E_\mu
a_{\mu}^{\dagger}a_{\mu} + \sum_i |{\bf h}_i| b^{\dagger}_{i}b_{i} \;,
\label{H_MF_bosonic}
\ee while corrections to the mean field the arise from the fluctuation
part of the interaction, Eq.~\eqref{eq:fluct}, neglected so far: \bea
&H_{\rm fluct}(i) \approx - G\; b^{\dagger}_{i}b_{i} : c_{i}^{\dagger
} {\bf e}_i^z {\bf F} c_{i} : \\ &\phantom{nnn}+ G\; \sqrt{\frac {S}
  2} \bigl( b^{\dagger}_{i} : c_{i}^{\dagger } {\bf e}_i^+ {\bf F}
c_{i} : + b_{i} : c_{i}^{\dagger } {\bf e}_i^- {\bf F} c_{i} :
\bigr)\;, \nonumber \eea where we defined ${\bf e}_i^\pm \equiv {\bf
  e}_i^x \pm i {\bf e}_i^y$, and $:\dots :$ denotes normal ordering
with respect to the mean field ground state. Thus the neglected
fluctuation terms just describe the interaction of spin waves with
spin excitations of the impurity band.

We can also rewrite the interaction part of the Hamiltonian using the
proper eigenstates of the mean field Hamiltonian and the corresponding
creation and annihilation operators as \bea H_{\rm fluct} &=& \sum_{i}
\sum_{\mu, \nu} \gamma_{\mu \nu}^{z} (i)\; b_i^{\dagger} b_i
:a_{\mu}^{\dagger} a_{\nu}:
\label{eq:h_imp}
\\ &+& \sum_{i} \sum_{\mu, \nu} \left ( \gamma_{\mu \nu}^{+} (i) \;
a_{\mu}^{\dagger} a_{\nu} b_i^{\dagger} + \gamma_{\mu \nu}^{-} (i) \;
a_{\mu}^{\dagger} a_{\nu} b_i \right )\; \nonumber \eea with the
couplings $\gamma_{\mu \nu}^{m}$ defined from Eq.~\eqref{eq:F_munu} as
\bea \gamma_{\mu\nu}^{z} &=& G \sqrt{\frac {S} 2}\,\, {\bf e}_i^z {\bf
  F}_{\mu\nu}(i)\;, \\ \gamma_{\mu\nu}^{\pm} &=& G \sqrt{\frac S
  2}\,\, {\bf e}_i^\pm {\bf F}_{\mu\nu}(i)\;.  \nonumber \eea

Having introduced a bosonic representation for the spins, we can now
use standard diagrammatic methods to compute spin-fluctation
corrections to the optical conductivity perturbatively in the small
coupling, $G$.  In the noninteracting theory, defined by
Eqs.~\eqref{eq:H_band} and \eqref{H_MF_bosonic}, we have three fields,
and, correspondingly, we have three Green's functions.  The
non-interacting Green's functions \bea G_{\rm imp}^{(0)}\left (\mu,
\omega_n \right ) = \frac 1 {i \omega_n -E_{\mu}}
\\ G^{(0)}_{\rm val}\left (k,j,\omega_n \right ) = \frac 1 {i \omega_n
  -\varepsilon_{k}(i)}
\eea describe single particle excitations in the impurity and valence
bands. We shall denote them by dashed and continuous lines,
respectively. Spin waves are described by the noninteracting bosonic
propagator,
\begin{equation}
D^{(0)}(j, \nu_n) =\frac 1 {i \nu_n - h_j} \;.
\end{equation}
This propagator is site-diagonal, and accounts for the spin precession
created by the local exchange field of the impurity band holes.  The
interaction part, Eq.~\eqref{eq:h_imp}, introduces then three types of
vertices between these Green's functions, which we depicted in
Table~\ref{table:green_functions}.

\begin{figure}[t]
\centering
\includegraphics[width=3.0in]{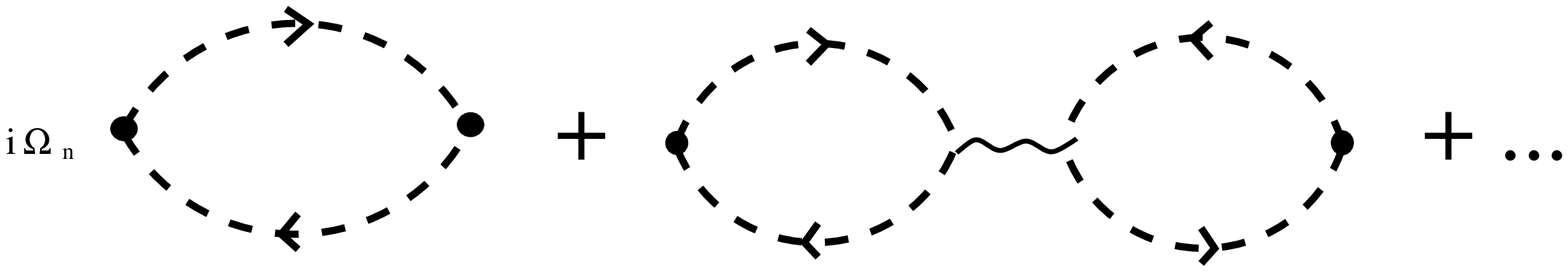} \vskip0.3cm
\includegraphics[width=2.5in]{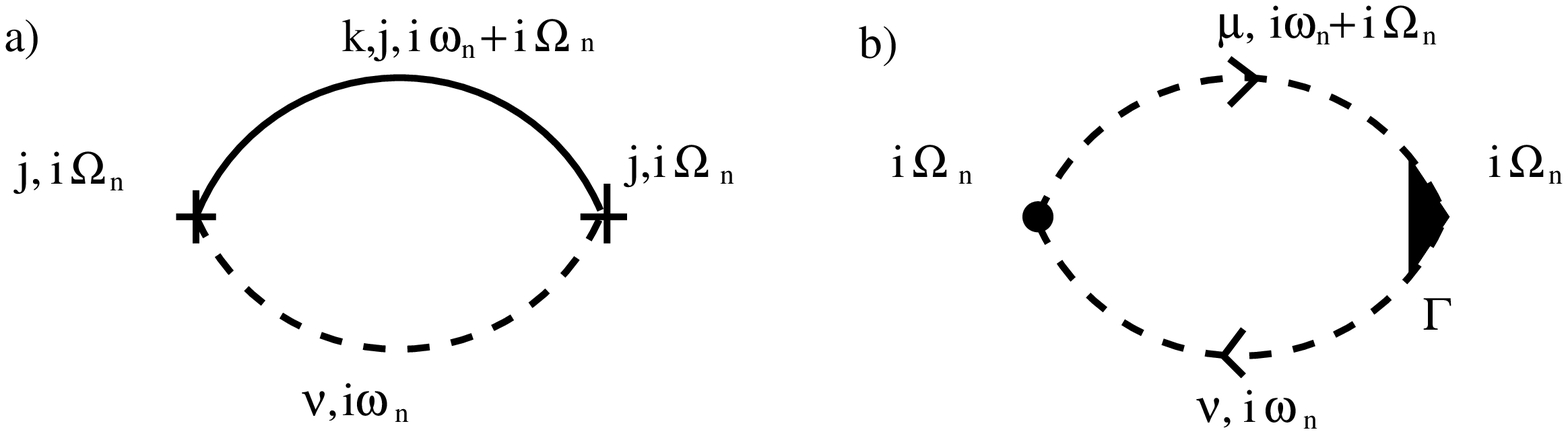}
\includegraphics[width=3.0in,clip]{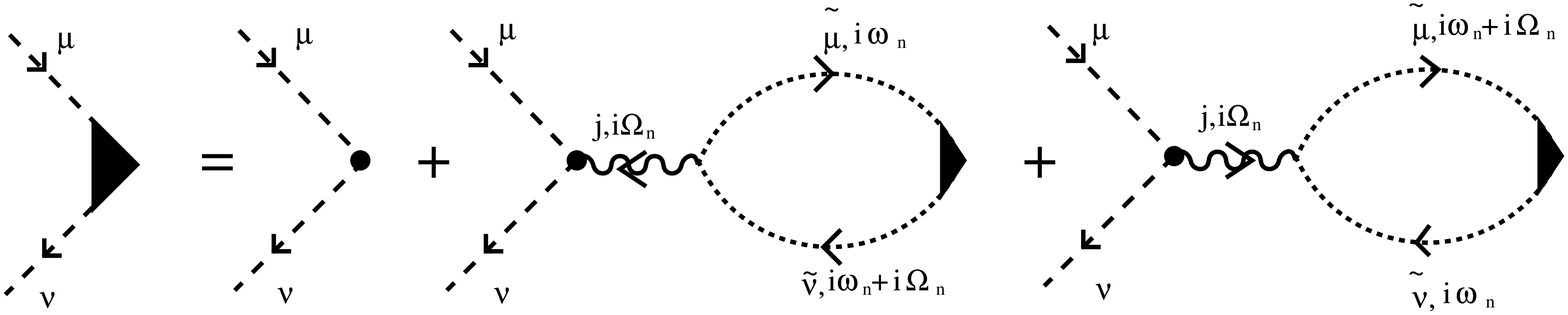}
\caption{ Top: RPA series giving the leading correction to the optical
  conductivity.  The dot represents the current vertex. Middle: (a)
  interband contribution to the optical conductivity and characterize
  excitations, (b) intraband contribution.  Bottom: Bethe-Salpeter
  equation for the renormalized current vertex $\tilde
  \Gamma_{\mu\nu}$, indicated as a filled .triangle.}
\label{fig:vertex_equation}
\label{fig:RPA}
\end{figure}
To compute the current-current response function, we shall use a
random phase approximation-type (RPA-type) approximation in the
bosonic line, which mediates the interaction between the valence band
holes: In the diagrammatic language this means that one neglects
self-energy corrections, and sums up only the diagrams shown in
Fig.~\ref{fig:RPA} (top layer). Notice that at $T=0$ the vertex
$\gamma^z$ does not give a contribution to the series in leading order
in $G$.  The RPA series can be converted into an integral equation
(Bethe-Salpeter equation), as shown in Fig.~\ref{fig:RPA} (bottom
layer), where one formally defines a renormalized current vertex
function, $\Gamma_{\mu\nu}$.  Expressed analytically, the vertex
equation reads

\begin{multline}
\sum_{\tilde \mu \tilde \nu}\left [ \delta_{\mu \tilde\mu} \delta_{\nu
    \tilde\nu} - \sum_{i} \gamma_{ \mu \nu}^{+} D ( -i\Omega_n,
  H_i)\Pi^0_{\tilde \mu,\tilde \nu}( i\Omega_n) \right .\\ \left
  .-\sum_{i} \gamma_{ \mu \nu}^{-} D ( i\Omega_n, H_i)\Pi^0_{\tilde
    \mu \tilde \nu}(i\Omega_n) \right ]\tilde \Gamma_{\tilde \mu
  \tilde \nu} = j_{\mu \nu}
\label{eq:BetheSalpeter}
\end{multline}
Here $\tilde \Gamma_{\mu \nu} (i \Omega_n)$  stands  for $\tilde
\Gamma_{\mu \nu} (i \Omega_n)= {\bf \Gamma}_{\mu \nu} (i \Omega_n){\bf
  e}_0$ with ${\bf e}_o$ the polarization of the electric field, and
similarly, $ j_{\mu \nu} = {\bf j}_{\mu \nu} {\bf e}_0$.  The
intraband polarization bubble is then expressed in terms of the bare
$j_{\mu \nu}$ and full vertex $\Gamma_{\mu \nu}$ as
\begin{multline}
\Pi^{\rm intra}_{JJ}(i \Omega_n) = \sum_{\mu, \nu} \; j_{ \mu \nu}
\Gamma_{\mu\nu}(i \Omega_n) \;\Pi^0_{\mu,\nu}(
i\Omega_n),,\label{eq:intraband}
\end{multline}
with the bare polarization buble defined as:
\begin{multline}
\Pi^0_{\mu,\nu}( i\Omega_n) = \;\sum_{i\omega_{n}} G_{\rm
  imp}^{(0)}(\mu, i\omega_n)\, G_{\rm imp}^{(0)}(\nu, i\omega_n+i
\Omega_n)\;
\end{multline}
Immediately, this gives for the intraband optical conductivity the
expression
\begin{multline}
\sigma_{intra}\left(\Omega\right) = \frac{i}{2\Omega} \sum_{\mu, \nu}
      { ({\bf j}_{\nu \mu}{\bf e}_0) ({\bf \Gamma}_{\mu \nu} (\Omega)
        {\bf e}_0)} \left [ f(E_{\mu})- f(E_{\nu})
        \right] \\ \times \left[
        \frac{1}{\Omega+E_{\mu}-E_{\nu}+i\delta} -
        \frac{1}{\Omega-E_{\mu}+E_{\nu}+i\delta}
        \right]\;.\label{eq:intra2}
\end{multline}
The mean field limit is recovered if one sets $\gamma\to0$ in the
Bethe-Salpeter equations, Eq.~\eqref{eq:BetheSalpeter}. Then
Eq.~\eqref{eq:intra2} reduces to Eq.~\eqref{eq:intra}.

We have solved the Bethe-Salpeter equations
Eq.~\eqref{eq:BetheSalpeter} numerically, and computed the resulting
corrections to the optical conductivity. However, we found that, at
typical optical frequencies, they do not give an important correction
to the optical conductivity as compared to the previously presented
mean field results. This is not very surprising: Magnetic excitations
have energies in the range of $T_C$, i.e. in the 5-6 meV range, which
is very small compared to the optical frequencies studied here. In the
frequency range $\omega< 5 {\rm meV}$ they are, however, expected to
result in interesting effects, and they may lead to some additional
features in the optical spectrum.

The present calculation has been carried out at $T=0$ temperature. In
this case the diagrammatic derivation of Eqs.~\eqref{eq:BetheSalpeter}
and \eqref{eq:intra2} is very transparent and straightforward.
However, we also derived these equations using the much more tedious
equation of motion method of Ref. [\onlinecite{Kassaian}], which holds also at finite temperatures. It
turns out that Eqs.~\eqref{eq:BetheSalpeter} and \eqref{eq:intra2} are
very general, and in the {\em same form}, they carry over to finite
temperatures too. The only modification is that the original mean
field Hamiltonian must be diagonalized using finite temperature
expectation values. Our formalism thus provides a way to study the
effects of finite temperature on the optical conductivity and the
interplay of ferromagnetism.  The study of these finite temperature
small frequency effects is, however, beyond the scope of the present
work.

\section{Conclusions}
\label{sec:conclusions}

In this paper, we presented a calculation of the optical properties of
\gamnas in the very dilute limit, where it can be described in terms
of an impurity band picture. Our approach consisted of constructing an
effective Hamiltonian with parameters determined from microscopic
variational calculations. The effective Hamiltonian obtained this way
not only accounts for the impurity band, but it also describes
transitions from the impurity band to the valence band.

Our mixed approach captures correctly the most essential features of
the experiments: It predicts a rather wide Drude peak, originating
from inter-impurity band transitions, which is partly merged with a
mid-infrared peak at $\sim 200 \;{\rm meV}$. The overall conductivity
values as well as the positions of these features are well-reproduced
by our theory. Remarkably, our calculations give a quantitative
description of the concentration-induced red-shift of the mid-infrared
peak, and gives optical mass and residual resistivity values that
agree well with the experimentally observed values.

Furthermore, we are also able to capture the metal-insulator
transition, which, depending on the level of compensation, occurs in
the range of $x= 0.1 - 0.3 \;\%$ for uncompensated samples while it is
in the range of $x= 1 \sim 2 \;\%$ for moderate compensations.  This
is in good agreement with the experiments. Our results thus
promote the theoretical picture that the metal-insulator transition
takes place \emph{within the impurity band}, well before it merges with the
valence band.\cite{Fiete, Berciu1} This picture is indeed supported by the
observation of a finite activation energy as one approaches the
metal-insulator transition.\cite{Blakemore} Further indirect evidence
for this scenario is given by Fig. 8 of Ref.~\onlinecite{Jungwirth3}:
at very small concentrations, where the impurity band is expected to
be insulating, and the conduction is due to activated behavior to the
valence band, the mobility of holes agrees with that of many other
alloys, that are known to have valence band conduction. However, right
after the metal-insulator transition ($x>x_C\approx 0.2\%$), the
mobility clearly drops to a much {\em smaller value}. The strong
deviation from other 'valence band' compounds could naturally be
explained by the fact that conduction is due to holes {\em within} the
impurity band, and that the metal-insulator transition also occurs
there, in agreement with our calculations.

We have also investigated the scaling of the conductivity and the
optical mass close to the metal-insulator transition. We find that the
intraband optical conductivity scales as $\sigma (\omega) = \sigma(0)+
A \omega^{1/2}$ on the metallic side, close to the
transition.\cite{Shapiro} Close to the transition point the optical
mass is strongly renormalized and takes values $\mopt \sim 10\; m_e$,
while far from the metal-insulator transition we find a mass in the
range $\mopt \approx m_e$.

\begin{figure}[t]
\centerline{\includegraphics[width=2.5in,clip]{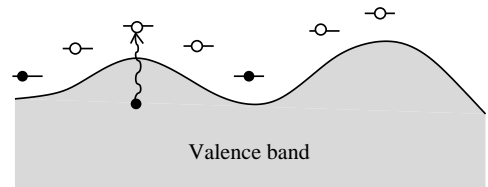}}\vspace*{3ex}
\caption{Deep, strongly localized occupied levels in the tail of the
  valence band move together with the valence band edge, and can give
  rise to impurity transitions at about {\rm 200\; meV}, indicated by
  the wavy line. {\rm Empty circles denote holes.}}
\label{fig:transition}
\end{figure}

Although our calculations were performed at the mean field level, we
extended it to incorporate spin wave fluctuations too. While the
formalism discussed here is only applicable at $T=0$ temperature, we
can also derive the same integral equations using an equation of
motion method (not discussed here) that carries over to finite
temperatures. The final equations presented here thus apply for finite
temperatures too. Nevertheless, we find that for typical optical
frequencies spin waves do not give a substantial contribution. They
may, however, influence the low frequency optical properties ($\hbar
\omega< 20 \;{\rm meV}$).

Our approach is designed to work in the limit of small concentrations,
and it should break down for large concentrations. Where exactly this
breakdown occurs, is not quite clear. In Ref.~[\onlinecite{Jungwirth3}]
it was argued that the break-down should appear at some active Mn
concentration in the range of $x\sim 1\%$.\footnote{For compensated
  \gamnas many of the substitutional Mn ions are believed to be
  inactive and not to participate in the ferromagnetic state due to
  the presence of interstitial Mn ions.}  However, surprisingly, all
spectroscopic data seem to favor a picture in terms of impurity band
physics even at intermediate
concentrations.\cite{Singley1,Singley2,Burch1, Okabayashi,STM,
  Linnarsson}  Also, 
our results are in very good agreement with the experimental optical
data, if we just blindly extrapolate them to $x\sim 3-4 \%$.
Theoretically, this is rather mysterious: One possible explanation
would be that  \gamnas is rather inhomogeneous, and some
spectroscopic data are dominated by regions of small Mn
concentration. The other possibility is that although the impurity
band density of states may merge with the valence band at larger
concentrations, most of the occupied states are in the tail of the
valence band, which has a strong impurity band character. This
scenario is sketched in Fig.~\ref{fig:transition}. Since optical
transitions are local for these states, and since the valence band and
the deep acceptor states are Coulomb shifted by approximately the same amount,
these levels deep in the gap give rise to an impurity state transition
at a frequency close to $\sim 200{\;\rm meV}$. This is indeed
supported by our calculations, where we find that the shifted local
density of states has a large peak coming from localized states in the
tail of the impurity band.  A third possibility would be that the
properties of \gamnas on its surface, tested by essentially all
spectroscopical probes, differ from those of the bulk, and thus the
effective concentration of Mn ions is somehow reduced there.

This research has been supported by Hungarian grants OTKA Nos. NF061726,
K73361, Romanian grant CNCSIS 2007/1/780 and CNCSIS ID672/2009, and by
NSERC and CIfAR Nanoelectronics.

\appendix
\section{Ground state wave function of a single ${\rm Mn}$ ion}
\label{app:1Mn}

In this appendix we present some details on the variational
calculation of the bound acceptor state of the Mn impurities and the
computation of the corresponding optical matrix elements.

\subsection{Acceptor wave function}

We describe the Mn acceptor by the following Hamiltonian ($\hbar
=m_{0}=e=1$),
\begin{equation}
H= -\frac{\gamma }{2}\nabla ^{2}-\frac{1}{\epsilon r} + V_{cc}(r)\;.
\label{eq:1Mn}
\end{equation}
Here the factor $\gamma =1.782$ describes the mass renormalization
term, and $\epsilon =12.65$ is the dielectric function for ${\rm GaAs}$. The
explicit form of the central cell correction, $V_{cc}(r)$ was given in
Eq.~\eqref{eq:V_cc}.

To compute the ground state wave function of \eqref{eq:1Mn}, we used
the following variational Ansatz
\begin{eqnarray}
\Psi (r) & = &\sum\limits_{i=1}^{n}A_{i}\psi (\alpha _{i},r)\;,
\\ \psi (\alpha ,r) & = & \expect{r|\alpha} = \frac{1}{\sqrt{\pi
}}\alpha ^{3/2}e^{-\alpha r}\;,
\label{eq:var_Psi}
\end{eqnarray}
In this expression we fixed the coefficients $\alpha ^{\prime }s$ and
used only the $A_{i}$'s as variational parameters. The variational
equation for the latter is simply a linear equation of the form:
\begin{equation}
\sum\limits_{j=1}^{n}H_{\alpha _{i}\alpha _{j}}A_{\alpha
  _{j}}=E\sum\limits_{j=1}^{n}S_{\alpha _{i}\alpha _{j}}A_{\alpha
  _{j}}.
\end{equation}
For the Hydrogenic wave functions used, the matrix elements $H_{\alpha
  \beta }=\langle \alpha| H|\beta\rangle $ as well as the overlap
parameters $S_{\alpha \beta } = \langle \alpha| \beta\rangle $ can be
computed analytically.

\begin{figure}[h]
\centerline{\includegraphics[width=3.0in,clip]{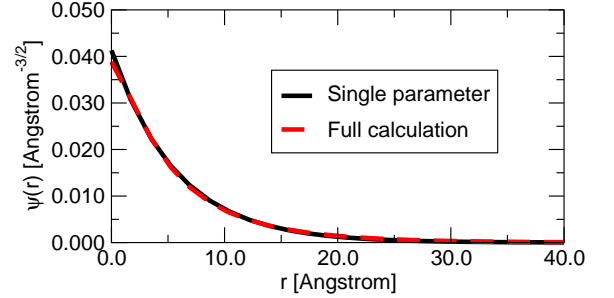}}\vspace*{3ex}
\caption{Normalized ground state wave function: Comparison between the
  single parameter and the full calculations.}
\label{wave_function_fig}
\end{figure}

The full variational solution computed using 50 parameters $\alpha_i$
is plotted in Fig.~\ref{wave_function_fig}, where it is also compared
to a single parameter variational solution.  The single-parameter
solution with $\alpha=0.091$ gives a remarkably accurate approximation
for the wave function, excepting the regime $r< r_0$, where small
deviations appear due to the central cell correction.

\subsection{Calculation of on-site optical matrix elements}
\label{section:onsite_optical}

Once the variational wave function \eqref{eq:var_Psi}  is known
(at hand), we can 
compute the optical matrix elements to valence band states. We have
two obvious choices: The simplest way to estimate these matrix
elements is to neglect the $1/r$ term as well as the central cell
correction (which does not interact too much with the $p$-channel), and
to use free electron states
\begin{equation}
\psi^{\rm free} _{lmk}(\mathbf{r})=\sqrt{2k^{2}}j_{l}(kr)Y_{lm}(\theta
,\varphi ),
\label{eq:free_scattering_states}
\end{equation}
with $j_{l}(kr)$ the spherical Bessel functions and $Y_{lm}(\theta
,\varphi )$ spherical functions. Alternatively, we can neglect only
the central cell correction and use Coulomb scattering states,
\begin{equation}
\psi^{\rm Coul}
_{lmk}(\mathbf{r})=\frac{\sqrt{2}}{r}F_{l}(kr)Y_{lm}(\theta ,\varphi
),
\label{eq:coulomb_scattering_states}
\end{equation}
where $F_l$ is related to the confluent hypergeometric function,
$_{1}F_{1}(a|b|x)$, as
\begin{equation}
F_{l}(kr)=c_{l}(\eta )e^{ikr}(kr)^{l+1}\quad_{1}F_{1}(l+1+i\eta
|2l+2|-2ikr),
\end{equation}
with $\eta =\frac{1}{\gamma \epsilon k}$, the constant $c_{l}(\eta
)=2^{l}e^{-\pi \eta /2}\frac{\left| \Gamma (l+1+i\eta )\right|
}{\Gamma (2l+2)}$, and $\Gamma (z)$ the gamma function.  Both
scattering states are normalized to satisfy $\langle lmk|l^{\prime
}m^{\prime }k^{\prime }\rangle = \pi \delta _{ll^{\prime}}\delta
_{mm^{\prime}}\delta (k-k^{\prime})$.

The matrix elements of $p_z$ can be computed analytically in both
cases.  For free valence electrons we obtain
\begin{equation}
\langle 1,0;k|p_{z}|\Psi \rangle_{\rm free}
=\frac{i}{\sqrt{3}}2^{5/2}\sum\limits_{i=1}^{n}A_{i} \frac{\alpha
_{i}^{5/2}k^{2}}{(\alpha _{i}^{2}+k^{2})^{2}}\;,
\label{eq:free_ME}
\end{equation}
while for Coulomb scattering states we have
\begin{eqnarray*}
&& \langle 1,0;k|p_{z}|\Psi \rangle_{\rm Coul}
  =\frac{i}{\sqrt{3}}2^{5/2}e^{-\pi \eta /2}\left| \Gamma (l+1+i\eta
  )\right| \\ &&\phantom{nn}\sum\limits_{j=1}^{n}A_{j}\frac{\alpha
    _{j}^{5/2}k^{2}}{
    _{j}-i k}{\alpha _{j}+ik}
\end{eqnarray*}
The matrix elements obtained this way were presented in
Fig.~\ref{fig:momentum_ground_state_coulomb}. The single parameter
ground state wave function gives a very good approximation in both
cases.  The overall energy dependence of the transition matrix
elements is qualitatively similar in Coulomb and the free scattering
state approximations. However, the height of the peak is about a
factor of two larger in the free electron approximation.

\section{Two  site problem}
\label{append:2Mn}

In this Appendix we determine the energies of the molecular orbitals
of an ${\rm Mn}_{2}$ system and use them to calculate the parameters
of an effective second quantized Hamiltonian.  We consider two Mn
ions, located at positions ${\bf r} = {\bf R}_{1,2} = \pm {\bf R}/2$,
and described by the Hamiltonian of Eq.~\eqref{eq:2site}. Similar to the
Mn ion case, we first construct the lowest lying 'molecular states'
using variational wave functions of the form,
\begin{equation}
\Psi (\mathbf{r})=\sum\limits_{\mu
  =s,p}\sum\limits_{i=1,2}\sum\limits_{l=1}^{N}A_{i,\;\mu ,l}\;\psi
_{i}^{(\mu )}(\alpha _{j},\mathbf{r}), \label{wave_function_eq}
\end{equation}
constructed from Hydrogen-like $s$ and $p$-type wave functions
centered at the Mn sites, $ {\bf R}_{1,2}$
\begin{eqnarray}
\psi_i^{(s)}(\alpha ,\mathbf{r})\equiv \langle {\bf
  r}|i,s,\alpha\rangle = \frac{1}{\sqrt{\pi }}\alpha ^{3/2}e^{-\alpha
  r_i}\;, \\ \psi_i ^{(p)}(\alpha ,\mathbf{r})\equiv \langle {\bf
  r}|i,p,\alpha\rangle = \frac{1}{4\sqrt{2\pi }}\alpha
^{5/2}x_ie^{-\alpha r_i/2}\;.
\end{eqnarray}
Here $\widehat{x}$ is the direction connecting the two sites, and
${\bf r}_i = {\bf r}- {\bf R}_i$ denote the position of the valence
hole relative to ${\bf R}_{1,2}$. We considered only states that do
not have a mirror plane that contains $\widehat{x}$, and therefore
included only $p_x$ orbitals. Similar to Appendix~\ref{app:1Mn}, we
fix the parameters $\alpha_j$ in Eq.~\eqref{wave_function_eq}, and
consider only the $A_{i,\;\mu,j}$ as variational parameters.

As in case of the single Mn ion problem, the coefficients
$A_{i,\;\mu,j}$ satisfy a set of linear equations, \be \sum_{\nu=s,p}
\sum_{j=1,2} \sum\limits_{l=1}^{N} \bigl[ H_{i,j}^{(\mu,\nu)}(\alpha_k
  ,\alpha_l ) - E \;S_{i,j}^{(\mu,\nu)}(\alpha_k ,\alpha_l )\bigr]
A_{j,\;\nu,l}\;,
\label{eigenvalue}
\ee with the overlap matrix elements and the Hamiltonian matrix
elements defined as
\begin{eqnarray*}
S_{i,j}^{(\mu ,\upsilon )}(\alpha ,\beta ) &=& \langle i,\mu,\alpha|
j,\nu, \beta\rangle \;,
\\ H_{i,j}^{(\mu,\nu)}(\alpha ,\beta ) &=& \langle i,\mu,\alpha| H |
j,\nu, \beta\rangle \;.
\end{eqnarray*}
The above integrals can be performed analytically using elliptic
coordinates. Some details of the derivation as well as the analytical
expressions are given as an on-line supplement.
 
\begin{figure}[t]
\centerline{\includegraphics[width=3.0in,clip]{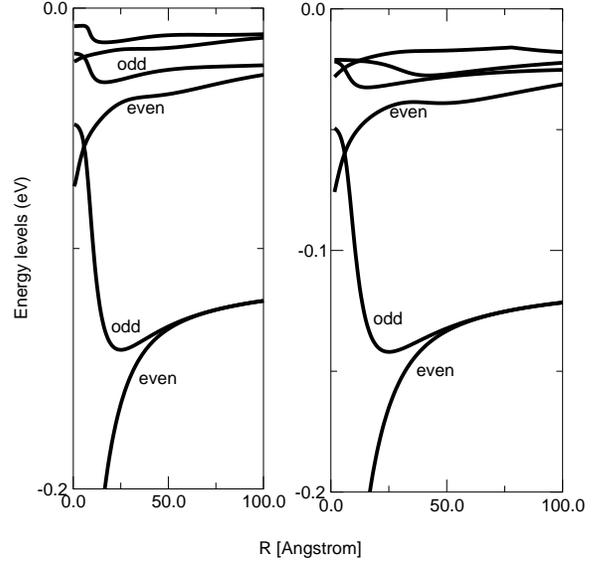}}\vspace*{3ex}
\caption{ The lowest few energy levels of the ${\rm Mn}_2$ molecule as
  a function of distance.  The left figure presents the case where
  only s-waves were considered (18 parameters were used). The right
  panel shows the results when both s and p waves are considered (14
  parameters were used). }
\label{energy_levels_fig}
\end{figure}

We remark here that Eq.~\eqref{eigenvalue} not only provides an
accurate estimate for the ground state energy, but it also accounts
for the first few excited states.  In Fig.~\ref{energy_levels_fig} we
show the first few lowest energy states obtained in this way as a
function of Mn separation, $R$. As demonstrated in the figure, the
inclusion of $p$-orbitals does not result in a major improvement for
the first five states, though it introduces a level crossing between
the first and the second excited states at a distance $R\sim 6 \AA $
where the tight-binding approximation fails.

The hopping $t$ and the energy $E$ appearing in Eq.~\eqref{2Mn_eff}
and shown in Fig~\ref{comparison_t_fig} can be extracted from this
spectrum as:
\begin{eqnarray}
&&t(R)=\frac{1}{2}(E_{\rm odd}(R)-E_{\rm even}(R)), \nonumber
  \\ &&E(R)=\frac{1}{2}(E_{\rm even}(R)+E_{\rm odd}(R)), \nonumber
\end{eqnarray}
where $E_{\rm even}$ and $E_{\rm odd}$ represent the lowest lying
(even) state of the ${\rm Mn}_2$ molecule and that of the first (odd)
excited state.

The electric field induces transitions between the
previously-mentioned even and odd states, and thereby generates a
coupling of the form, Eq.~\eqref{eq:H_ext}.  To determine the
parameters appearing in this equation, one simply needs to compute the
momentum matrix elements $\langle \psi _{\rm odd}(r)|{\bf p}|\psi
_{\rm even}(r)\rangle$ that enter the coupling to the external
electromagnetic field.  Since we aligned the two $\rm{Mn}$ ions along
the $x$ direction the only non-vanishing component of the momentum
matrix elements is $\langle \psi _{\rm odd}(r)|{ p}_x|\psi _{\rm
  even}(r)\rangle$. It is possible to express this matrix element
analytically for our variational wave functions, though the final
expressions are rather cumbersome.  The final results have
been plotted in
Fig.~\ref{peierls_substitution_single_parameter_full_calculation_fig}.





\end{document}